\documentclass[a4paper,10pt]{JHEP3}
\usepackage[dvips]{graphicx}
\usepackage{cite}
\usepackage{amssymb}
\usepackage{amsmath}
\usepackage{latexsym}
\usepackage{xspace}
\usepackage{array}
\usepackage{url}
\usepackage{afterpage}

\usepackage{tabularx}
\newcolumntype{L}[1]{>{\raggedright\arraybackslash}p{#1}} 
\newcolumntype{C}[1]{>{\centering\arraybackslash}p{#1}}

\newcommand{\trule}{\rule[-1.5mm]{0mm}{6mm}}
\newcommand{\SqrtS}{\sqrt{S}}
\newcommand{\alphas}{\alpha_s}

\newcommand {\uL}{\tilde{u}_{L}}
\newcommand {\uR}{\tilde{u}_{R}}
\newcommand {\dL}{\tilde{d}_{L}}
\newcommand {\dR}{\tilde{d}_{R}}

\newcommand{\sq}{\ensuremath{\tilde{q}\xspace}}
\newcommand{\slep}{\ensuremath{\tilde{l}\xspace}}

\newcommand{\msq}{\ensuremath{m_{\sq}}}

\newcommand{\neu}{\ensuremath{\tilde{\chi}^{0}}}

\newcommand{\neuone}{\ensuremath{\tilde{\chi}^{0}_1}\xspace}
\newcommand{\neutwo}{\ensuremath{\tilde{\chi}^{0}_2}\xspace}
\newcommand{\LSP}{\neuone}
\newcommand{\gluino}{\ensuremath{\tilde{g}}}
\newcommand{\gl}{\ensuremath{\tilde{g}}\xspace}

\newcommand{\MeV}{{\rm Me\kern -1pt V}\xspace}
\newcommand{\GeV}{{\rm Ge\kern -1pt V}\xspace}
\newcommand{\TeV}{{\rm Te\kern -1pt V}\xspace}

\newcommand{\qLchainX}{\ensuremath{\sq_L\to& q~\neutwo &\to q~l^{\pm}~\slep_{L/R}^{\mp} \to q~l^{\pm}~l^{\mp}~\neuone\xspace}} 
\newcommand{\qRchainX}{\ensuremath{\sq_R\to& q~\neuone&}}
\newcommand{\qLRchain}{\ensuremath{\sq_L\to q~\neutwo \to q~l^{\pm}~\slep_{R}^{\mp} \to q~l^{\pm}~l^{\mp}~\neuone\xspace}}
\newcommand{\qLLchain}{\ensuremath{\sq_L\to q~\neutwo \to q~l^{\pm}~\slep_{L}^{\mp} \to q~l^{\pm}~l^{\mp}~\neuone\xspace}}

\newcommand{\msbar}{\ensuremath{\overline{\text{MS}\xspace}}}

\newcommand{\be}{\begin{eqnarray*}}
\newcommand{\ee}{\end{eqnarray*}}
\newcommand{\bee}{\begin{eqnarray}}
\newcommand{\eee}{\end{eqnarray}}
\renewcommand{\eqref}[1]{eq.~(\ref{#1})}

\newcommand{\figref}[1]{figure~\ref{#1}}
\newcommand{\tabref}[1]{table~\ref{#1}}

\newcommand{\fba}{\unskip\,{fb}\xspace}

\usepackage{cancel}
\usepackage{dcolumn}
\newcolumntype{d}[0]{D{.}{.}{-1}}

\newcommand{\eg}[0]{e.g.}

\newcommand{\softsusy}[0]{\texttt{SOFTSUSY}\xspace}
\newcommand{\sdecay}[0]{\texttt{SDECAY}\xspace}

\newcommand{\fastjet}[0]{\texttt{FastJet~3.0.2}\xspace}

\def\missingET{\ensuremath{\displaystyle{\not}E_T}\xspace}

\newcommand{\mjln}{\ensuremath{m_{jl_{n}}\xspace}}
\newcommand{\mjlf}{\ensuremath{m_{jl_{f}}\xspace}}
\newcommand{\mjlm}{\ensuremath{m_{jl^{-}}\xspace}}
\newcommand{\mjlp}{\ensuremath{m_{jl^{+}}\xspace}}
\newcommand{\mjll}{\ensuremath{m_{jll}\xspace}}
\newcommand{\mjlhigh}{\ensuremath{m_{jl(\text{high})}\xspace}}
\newcommand{\mjllow}{\ensuremath{m_{jl(\text{low})}\xspace}}
\newcommand{\mjlltresh}{\ensuremath{m_{jll(\text{thresh})}\xspace}}
\newcommand{\mjlu}{\ensuremath{m_{jl(\text{u})}\xspace}}
\newcommand{\mjld}{\ensuremath{m_{jl(\text{d})}\xspace}}
\newcommand{\mjls}{\ensuremath{m_{jl(\text{s})}\xspace}}
\newcommand{\mjlpp}{\ensuremath{m_{jl(\text{p})}\xspace}}

\title{On cascade decays of squarks at the LHC \\ in NLO QCD}

\author{W. Hollik, J. M. Lindert, D. Pagani\\
 Max-Planck-Institut f\"ur Physik, 
 F\"ohringer Ring 6, 
 D-80805 M\"unchen, Germany\\
 Email: \email{hollik@mpp.mpg.de}, \email{lindert@mpp.mpg.de}, \email{pagani@mpp.mpg.de}}

\abstract{
In this paper we present an analysis at NLO of the contribution from squark-squark production to the experimental signature $2j+l^{+}l^{-}+\missingET(+X)$ with opposite-sign same flavor leptons, taking into account decays and experimental cuts. We consider the case in which one squark decays directly into the lightest neutralino $\neu_{1}$ and the other one into the second lightest neutralino and subsequently into $l^{+}l^{-}\neu_{1}$ via an intermediate slepton. On one hand we study effects of the NLO corrections on invariant mass distributions which can be used for future parameter determination. On the other hand we analyze the impact on predictions for cut-and-count searches using this experimental signature. 
}

\preprint{MPP-2013-38}

\begin{document}

\section{Introduction}
\label{sec:intro}

Supersymmetry (SUSY) remains one of the theoretically most appealing models for physics beyond the standard model (SM), and the light Higgs boson predicted by the Minimal Supersymmetric Standard Model (MSSM) is consistent with the recent observation of a Higgs-like bosonic resonance \cite{:2012gu,:2012gk}. 
However, for a natural stabilization of the EW sector, direct signs of SUSY should hopefully be observable at the \TeV scale.
Until now there has not been any direct evidence for deviations from the SM and the experiments at the LHC have set new limits in various regions of the parameter space of the MSSM, investigating numerous final states and signatures \cite{Chatrchyan:2012te,Chatrchyan:2012gq,Chatrchyan:2013wc,Aad:2011xk,Aad:2012rz}.

Once a clear signal for physics beyond the Standard Model (BSM) is established, the character and parameters of the underlying model have to be determined. Only a precise and non-ambiguous determination of \TeV scale parameters allows reconstruction of the underlying theory \cite{ArkaniHamed:2005px,Bornhauser:2012iy} and, $\eg$ in case of SUSY, investigation of different breaking scenarios. Moreover, different BSM models result in similar signatures at the LHC, so even the determination of the general model would be a major challenge \cite{Hubisz:2008gg}. In recent years various techniques for these challenges have been developed. Many of these techniques rely on the occurrence of cascade decay chains where, due to an additional symmetry, the lightest new particle (LSP in the SUSY case) is stable and, motivated by the possibility to provide a dark matter candidate, unobservable in the detectors. For a review of mass determination techniques see, \eg,  \cite{Barr:2010zj}.\\

In this paper we investigate NLO (S)QCD corrections to the ``qll-chain'', also known as ``golden decay chain'',
\bee
\label{qLchain}
\qLchainX\, ,
\eee
where a left-handed squark decays into a quark and a second lightest neutralino \neutwo, which subsequently decays via an intermediate slepton into a pair of opposite-sign same-flavour (OS-SF) leptons and a lightest neutralino \neuone. For a systematic treatment we combine this decay chain  at NLO with the production of a pair of squarks $\sq^{\phantom{'}}_L\sq_R'$, where the second squark decays directly into a quark and a \neuone,
\bee
\label{qRchain}
\qRchainX \, .
\eee
Thus we provide, for the first time, a fully differential description at NLO QCD of the contribution to the signature $2j+l^{+}l^{-} (\text{OS-SF}) + \missingET(+X)$ from the process
\bee
\label{eq:process}
 pp\rightarrow \sq^{\vphantom{'}}_{L}\sq'_{R}\rightarrow q \neu_{1}q'l^{\pm}l^{\mp}\neu_{1} \, .
\eee
For heavy squarks and gluinos, as suggested by LHC exclusion limits,  the considered squark--squark production is the dominant process among all coloured SUSY production channels, see for example  \cite{Falgari:2012hx}. 
In our analysis we assume $\neu_{1}$ and $\neu_{2}$ to be mainly bino- and wino-like, as they appear in large parameter regions of models with unified gaugino masses at the GUT scale. In such scenarios the other decays, $\sq_{L}\rightarrow\neu_{1}q$ and $\sq_{R}\rightarrow\neu_{2}q$, are highly suppressed for squarks of the first and second generation.
Furthermore, in the benchmark scenarios we consider, all relevant squarks are lighter than the gluino \gl, so squarks can exclusively decay into neutralinos and charginos.  \\

The decay chain, \eqref{qLchain}, was introduced in \cite{Hinchliffe:1996iu,Bachacou:1999zb} and studied in many subsequent works
\cite{Allanach:2000kt,Weiglein:2004hn,Gjelsten:2004ki,Miller:2005zp,Lester:2006yw,Horsky:2008yi,Bisset:2008hm,Costanzo:2009mq,Polesello:2009rn,Matchev:2009ad,Matchev:2009iw,Cheng:2009fw,Edelhauser:2010gb,Agashe:2010gt,Nojiri:2010mk,Barr:2011xt,Chen:2011cya,Choi:2011ys}. 
%
These analyses showed that measurements of resulting invariant mass distribution endpoints and shapes can be exploited to determine the masses of the intermediate SUSY particles. Shapes of various relevant invariant mass distributions have first been calculated analytically at LO in \cite{Miller:2005zp} and they might be important to resolve ambiguities in mass measurements from kinematic endpoints \cite{Gjelsten:2006tg}. In \figref{fig:LO_shapes}, for illustrative purposes, normalized invariant mass distributions in $\mjll$, $\mjlhigh$ and $\mjllow$ are displayed for two benchmark points 
(with details given in the corresponding sections of this paper).
At this point we only want to emphasize that such shapes are very sensitive to the model parameters; on the other hand,
they can be distorted by higher order corrections, affecting thus the accuracy of endpoint determination techniques. 
%
\FIGURE{
 \includegraphics[width=.30\textwidth,angle=-90]{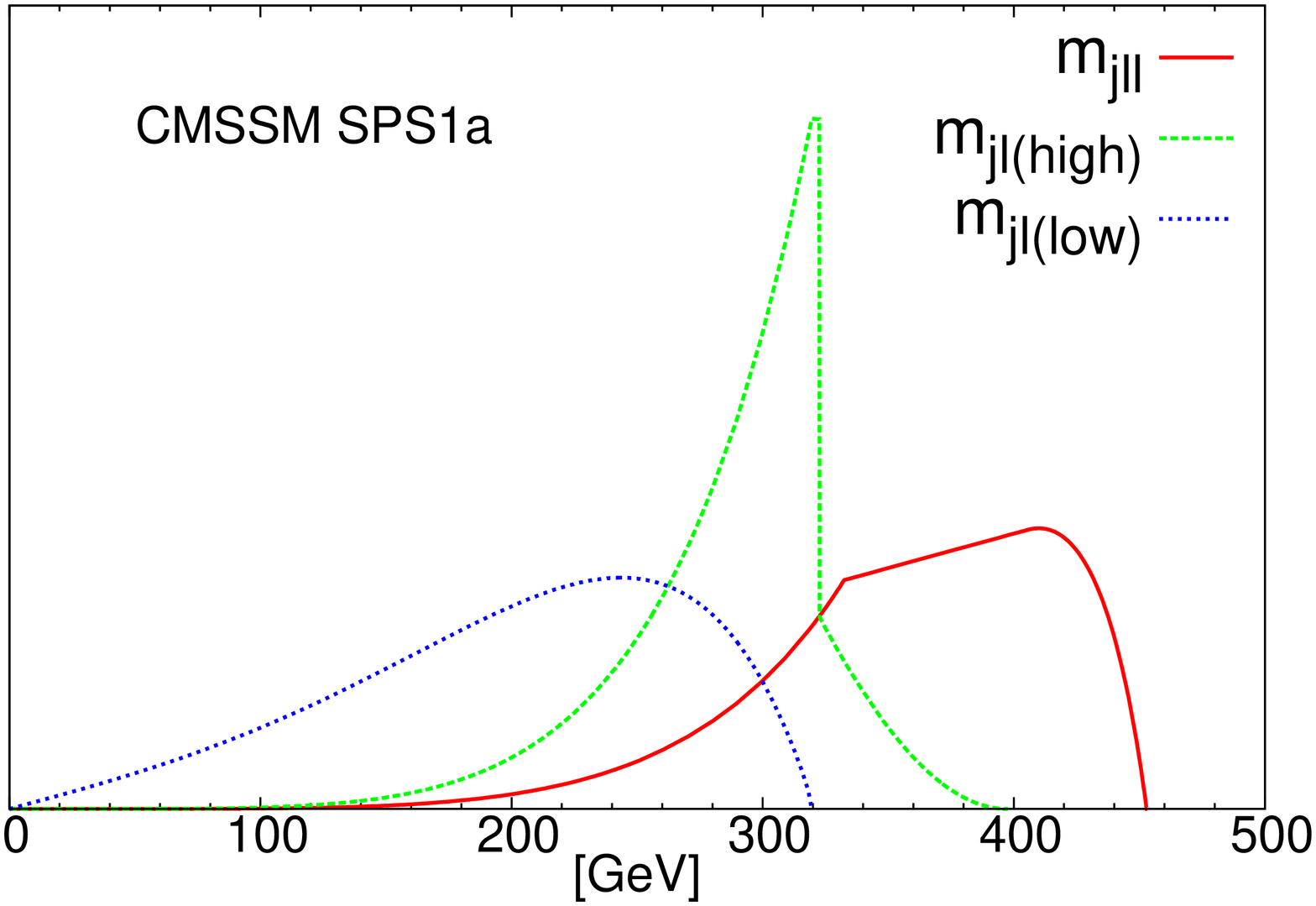}
 \includegraphics[width=.30\textwidth,angle=-90]{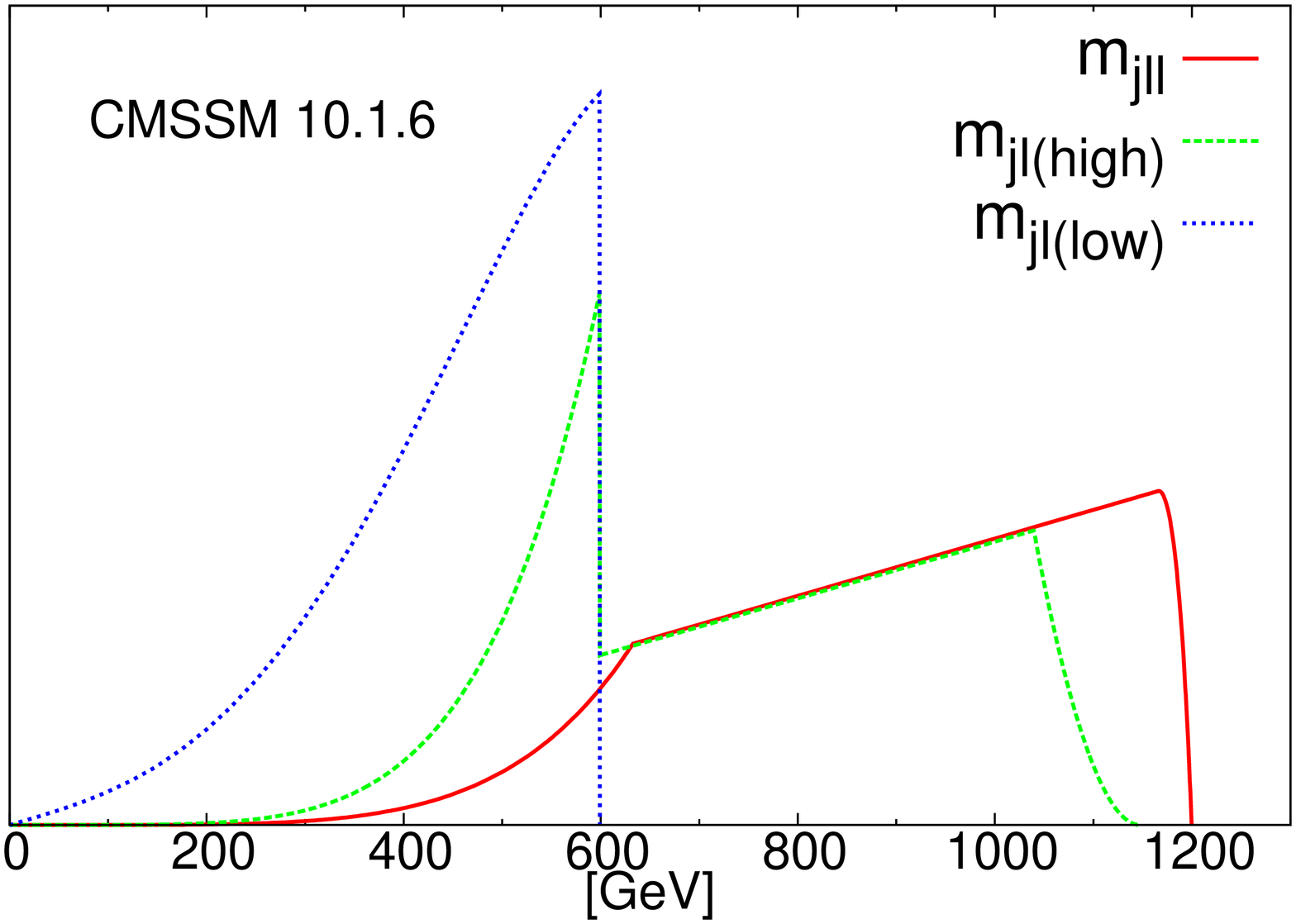}
\caption{Shapes of the $\mjll$, $\mjlhigh$ and $\mjllow$ distributions for SPS1a and 10.1.6 at LO.}
\label{fig:LO_shapes}
}

Such invariant mass distributions, due to correlations between particles in the decay chain, can be used also for spin measurements of the intermediate sparticles~\cite{Barr:2004ze}. 
In this way, \eg, a SUSY model can be discriminated from an Universal Extra Dimension (UED) model, see \eg, \cite{Smillie:2005ar,Athanasiou:2006ef} and \cite{Battaglia:2005ma,Datta:2005zs,Wang:2006hk,Choi:2006mr,Kilic:2007zk,
Wang:2008sw,Choi:2008pi,Burns:2008cp,Gedalia:2009ym,Ehrenfeld:2009rt,Srimanobhas:2011zz}. 
Additionally, besides observables based on invariant mass distributions, the signature $2j+l^{+}l^{-} (\text{OS-SF}) + \missingET(+X)$ can be used for SUSY searches at the LHC. Basic kinematical cuts can reduce SM backgrounds significantly and already now stringent bounds on relevant parameter regions have been obtained \cite{Chatrchyan:2012te, Aad:2011xk}. Furthermore, resulting rates can also be used for parameter determination within a global fit \cite{Dreiner:2010gv}. \\

In all these analyses a detailed understanding of theoretical uncertainties and effects from higher-order contributions is necessary, requiring precise predictions at a fully differential level including NLO corrections in production and decay.
In \cite{Horsky:2008yi} the ``qll-chain'' was investigated at NLO QCD in the squark rest frame.  There, real gluon radiation contributions are given in a fully analytical form and leading soft- and collinear gluon contributions are resummed. Furthermore, an LHC analysis using LO production matrix elements is presented. In their numerical analysis the authors 
of ~\cite{Horsky:2008yi} concentrate on the effect on distributions sensitive to spin correlations. 
In the present paper we continue the analysis of higher-order corrections of the ``qll-chain'' by combining the
production of squark pairs and their subsequent decays at the NLO level in a consistent way.
This allows a systematic study of the important jet combinatorics issues in the presence of
more than one final-state jet, emerging
from both real radiation at NLO and the 
combination of decay and production \cite{Nojiri:2010mk}. 
Futhermore, we investigate, besides the effects on spin determination,  the impact of such corrections on distributions important for mass determination and on inclusive event rates after experimental cuts.

Many details of the calculation, not explicitly discussed here, can be found in \cite{Hollik:2012rc} where an analogue analysis at NLO QCD of production of squark pairs directly decaying into two lightest neutralinos has been performed. Also another similar calculation for stop production and direct decay was recently performed in \cite{Boughezal:2012zb}. Conversely, calculations of higher-order contributions to either production
\cite{Beenakker:1994an,Beenakker:1995fp,Beenakker:1996ch,Beenakker:1997ut,Langenfeld:2009eg,Kulesza:2008jb,Kulesza:2009kq,Beenakker:2009ha, Beenakker:2010nq,Beneke:2010da,Beenakker:2011sf,Falgari:2012hx,Bornhauser:2007bf,Arhrib:2009sb,Hollik:2007wf,Hollik:2008yi,Hollik:2008vm,Beccaria:2008mi,Mirabella:2009ap,
Germer:2010vn,Germer:2011an,Falgari:2012sq,GoncalvesNetto:2012yt}
or  decay 
\cite{Hikasa:1995bw,Djouadi:1996wt,Kraml:1996kz,Bartl:1994bu,Bartl:1998xk,
Beenakker:1996dw,Beenakker:1996de,Guasch:1998as,Guasch:2002ez,Arhrib:2004tj,Arhrib:2005ea,Li:2002ey,Weber:2007id} 
of squarks (and gluinos) are manifold and include electroweak NLO corrections as well as  QCD corrections beyond NLO. 
The study of the signature $2j+l^{+}l^{-} (\text{OS-SF}) + \missingET(+X)$ presented here includes the contribution from squark--squark production, which constitutes the dominant production channel 
in the case of heavy squarks and gluinos, as already pointed out.
This work should be understood as a first step towards a calculation including all channels and also off-shell and non-factorizable corrections.\\

The outline of this paper is the following. In section \ref{sec:method} we briefly explain the method, based on the previous study \cite{Hollik:2012rc}, to combine production and decay at NLO. In section \ref{sec:decay_chain} we describe the calculation of the NLO corrections to just the decay chain, \eqref{qLchain}. In section \ref{sec:numerics} we present the numerical results.  This section \ref{sec:numerics} is dived in three subsections: the discussion of the considered benchmark scenarios, the analysis of NLO corrections to the isolated decay chain, and finally the discussion, for combined production and decay, of NLO effects on invariant mass distributions and also on event rates after kinematical cuts.  Conclusions and a possible outlook are given in \ref{sec:conlusion}.

%
\section{Method}
\label{sec:method}

We investigate the production of $\sq_L\sq'_R~(\sq^{*}_L\sq'^{*}_R)$ pairs induced by proton-proton collisions,
where the right-handed squark directly decays into the lightest neutralino $\neu_{1}$ and the left-handed one into a $\neu_{2}$ and subsequently into $l^{+}l^{-}\neu_{1}$ via an intermediate slepton $\slep^{\pm}_{L/R}$. This process results in the signature $2j+l^{+}l^{-}+\missingET(+X)$ and is illustrated in \figref{fig:blobs}. 
\FIGURE{
\includegraphics[width=0.65\textwidth]{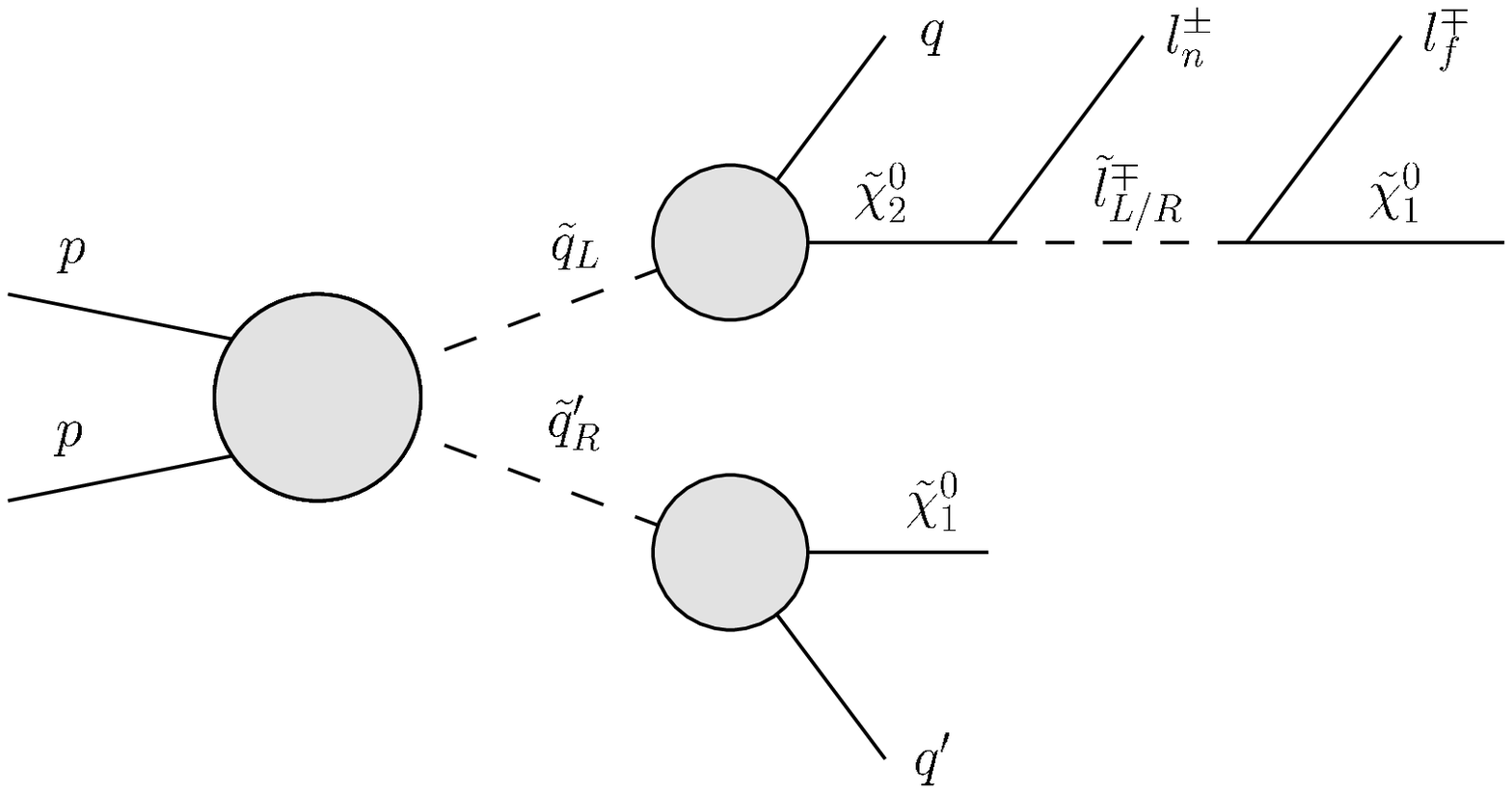}
\caption{General structure of factorizable NLO QCD corrections to the given process.\label{fig:blobs}}
}

At LO the only partonic subprocess that contributes to a given intermediate $\sq_{L}\sq'_{R}$  $(\sq^{*}_{L}\sq'^{*}_{R})$ configuration arises from a quark (anti-quark) pair $qq'$ $(\bar{q}\bar{q}')$ in the initial state. 
In our calculation we include contributions from all (s)quark flavors of the first two generations.
We will perform the following discussion without referring to the
charge-conjugate subprocesses; in the final results, however,  we
include the charge-conjugate processes explicitly.
Indeed, due to chirality-dependent interactions, the decay of an
anti-squark has to be treated independently 
from the corresponding squark decay.

Using the same theoretical framework as applied in
\cite{Hollik:2012rc} for squark-squark production and direct decays
into lightest neutralinos, we include the NLO factorizable corrections in
the narrow-width-approximation (NWA) to the process defined in
\eqref{eq:process}. 
They correspond, as illustrated in \figref{fig:blobs}, to the separate sets of 
corrections for the production and for the decays with squarks treated on-shell. 
For any flavor configuration, a systematic expansion of the
differential cross section in 
the strong coupling $\alpha_{s}$ yields
\begin{align}
\label{masterformel}
d\sigma^{(0+1)}_{\mbox{NWA}}(pp&\rightarrow\sq_{L}\sq'_{R} \rightarrow q\neu_{1}q'l^{+}l^{-}\neu_{1}(+X))\,=\\
&\frac{1}{\Gamma^{(0)}_{\sq_{L}}\,\Gamma^{(0)}_{\sq'_{R}}} \,
\Big[d\sigma^{(0)}_{pp\rightarrow\sq_{L}\sq'_{R} }\, 
     d\Gamma^{(0)}_{\sq_{L} \to q\neu_{1}l^{+}l^{-}} \,d\Gamma^{(0)}_{\sq'_{R} \to q'\neu_{1}}\,
     \Big(1-\frac{\Gamma^{(1)}_{\sq_{L}}}{\Gamma^{(0)}_{\sq_{L}}}-
            \frac{\Gamma^{(1)}_{\sq'_{R}}}{\Gamma^{(0)}_{\sq'_{R}}}\Big) \nonumber\\[0.2cm]
+&\, d\sigma^{(0)}_{pp\rightarrow\sq_{L}\sq'_{R} }\, 
     d\Gamma^{(1)}_{\sq_{L} \to q\neu_{1}l^{+}l^{-}}\, d\Gamma^{(0)}_{\sq'_{R} \to q' \neu_{1}}
\,+\,d\sigma^{(0)}_{pp\rightarrow\sq_{L}\sq'_{R} }\, d\Gamma^{(0)}_{\sq_{L} \to q\neu_{1}l^{+}l^{-}}\,
     d\Gamma^{(1)}_{\sq'_{R} \to q' \neu_{1}} \nonumber\\[0.2cm]
+&\, d{\sigma}^{(1)}_{pp\rightarrow\sq_{L}\sq'_{R}}\, d\Gamma^{(0)}_{\sq_{L} \to q\neu_{1}l^{+}l^{-}}\,
     d\Gamma^{(0)}_{\sq'_{R} \to q' \neu_{1}}\Big]  \, .\nonumber
\end{align}
In \eqref{masterformel},  $\Gamma^{(0)}_{\sq_{L}}$ and $\Gamma^{(0)}_{\sq'_{R}}$ denote the LO total
widths of the two squarks. The terms $d\Gamma^{(0)}_{\sq_{L} \to q \neu_{1}l^{+}l^{-}}$ and
$d\Gamma^{(0)}_{\sq'_{R} \to q' \neu_{1}}$ 
are the LO differential distributions for the decay chain and for the direct decay into the lightest neutralino,
respectively, boosted to the moving frames of $\sq_{L}$ and $\sq'_{R}$.
The other ingredient, 
$d\sigma^{(0)}_{pp\rightarrow\sq_{L}\sq'_{R}}$, 
is the LO hadronic differential production cross section.
The corresponding NLO corrections to the aforementioned quantities are indicated by an apex $(1)$.
The second line of \eqref{masterformel} contains a 
global factor with NLO contributions to the decays widths;
the third line includes, as the first term, the corrections to the
decay chain and, as second term, the corrections to the decay into the lightest neutralino;
the last line represents the corrections to the squark--squark production cross section.
The terms in \eqref{masterformel} refering to the LO and NLO contributions to squark--squark
production and to the direct squark decays into the lightest neutralino  
were calculated and discussed in~\cite{Hollik:2012rc}; the treatment
of the decay chain is addressed in section~\ref{sec:decay_chain}. On the basis of \eqref{masterformel} intermediate events for production and decay are combined in analogy to the strategy explained in \cite{Hollik:2012rc}.

%
\section{Calculation of the squark decay chain}
\label{sec:decay_chain}

The structure of the decay chain is illustrated in \figref{fig:decay}. With $l^{\pm}_{n}$ and $l^{\mp}_{f}$ we indicate, respectively, the lepton emerging from the  $\neu_{2}$ decay, the \textit{near lepton}, and the lepton emerging from the slepton decay, the \textit{far lepton}. Experimentally these leptons are indistinguishable.
\FIGURE{
\includegraphics[width=0.45\textwidth]{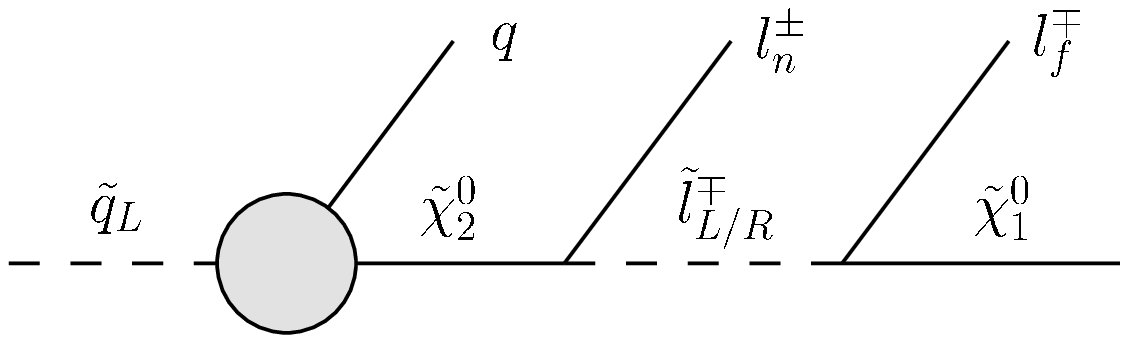}
\caption{Structure of the decay chain. NLO QCD corrections involve only the first step $\sq_{L}\rightarrow q \neu_{2}$.
}
\label{fig:decay}
} 
Figure~\ref{fig:decay} represents, in a compact notation, the four Feynman diagrams contributing to the tree level amplitude for the decay $\sq_{L}\rightarrow ql^{+}l^{-}\neu_{1}$. They correspond to the two cases $\sq_{L}\rightarrow ql_{n}^{+}l^{-}_{f}\neu_{1}$ and $\sq_{L}\rightarrow ql_{n}^{-}l_{f}^{+}\neu_{1}$ with a left- or right-handed intermediate (anti)slepton.

For non-degenerate left- and right-handed sleptons, as in the scenarios investigated in this paper, in NWA the structure of the squared amplitude of this decay chain becomes much simpler. In the limit $\frac{\Gamma}{m}\rightarrow 0$ for the sleptons and for $\neu_{2}$, the interferences between different diagrams vanish and the phase-space of the intermediate particles can be treated on-shell.
In this way LO and NLO contributions to the differential distribution for the decay chain in Figure~\ref{fig:decay} 
can be written as follows,
\begin{align}\label{intonf}
d\Gamma^{(0,1)}_{\sq_{L}\rightarrow q l^{+}l^{-}\neu_{1}}= \sum_{\sigma=\pm 1/2}
\frac{d\Gamma^{(0,1)}_{\sq_{L}\rightarrow q \neu_{2,\sigma}}} {\Gamma_{\neu_{2}}} \;
[d\Gamma_{\neu_{2,\sigma}\rightarrow l^{+}_{n}l^{-}_{f}\neu_{1}}+d\Gamma_{\neu_{2,\sigma}\rightarrow l^{-}_{n}l^{+}_{f}\neu_{1}}]~.
\end{align}
The index $\sigma$ represents the helicity of $\neu_{2}$. We keep, at LO and NLO, the full helicity dependence. 
Technically, we use, both at LO and at NLO, the matrix elements for the entire chain and set consistently the different on-shell conditions 
according to the intermediate states for the various contributions. In this way, 
the sum over the helicity states is automatically performed and off-shell effects can be switched on easily in a further study. The two terms contained in the square brackets of \eqref{intonf} correspond to the two different charge configurations for near and far leptons in the decay of $\neu_{2}$. Both configurations get contributions from the left- and the right-handed slepton.

As mentioned above, we concentrate on scenarios where the $\neu_{2}$ is dominantly wino-like. Thus, in the considered decay chain the coupling to a right-handed slepton, $\slep_R$, is heavily suppressed compared to the corresponding decay chain via a $\slep_L$. Only if the decay into a left-handed slepton is kinematically forbidden, $m_{\slep_{L}}>m_{\neu_{2}}>m_{\slep_{R}}$, the decay via a $\slep_R$ can contribute substantially.
Basically also the decay  of the $\tilde{\chi}_2^0$ into a OS-SF lepton pair via a $Z$ boson contributes when kinematically allowed; these effects are, however, numerically not significant for the benchmark points considered in this paper.

The calculation of $d\Gamma^{(1)}_{\sq_{L}\rightarrow q \neu_{2,\sigma}}$ is performed following the  procedure explained in~\cite{Hollik:2012rc}. 
$d\Gamma^{(1)}_{\sq_{L}\rightarrow q \neu_{2,\sigma}}$ includes contributions from loop corrections and from real gluon radiation. Consequently, in collinear and infrared safe regions, our NLO corrections are fully differential in the momentum of the gluon emitted. The rest of the electroweak decay chain is not affected by NLO QCD corrections.

%
\section{Phenomenological results}
\label{sec:numerics}
In this section we present the numerical results of our calculation. First, we
specify input parameters and relevant observables. Second, NLO corrections
to the decay chain, \eqref{qLchain}, not combined with the production, are investigated in the squark rest frame. 
Third, we present results for the decay chain combined, according to section \ref{sec:method}, with $\sq_L\sq'_R$ production.

\subsection{Parameters and observables}
Standard Model input parameters are chosen according to \cite{Nakamura:2010zzi}.
As PDF set we use CTEQ6.6~\cite{Nadolsky:2008zw} with the associated $\alphas^{\msbar}(\mu_R)$ at NLO. Renormalization scale $\mu_R$ and factorization scale $\mu_F$  
are both set to the average mass of all light-flavor squarks, 
$\mu=\mu_F=\mu_R=\overline{\msq}$. \\
Numerical results in this paper are presented for two representative benchmark scenarios and the LHC with a center of mass energy of $\SqrtS=14~\TeV$.   
We choose the two CMSSM scenarios SPS1a and 10.1.6 defined in \cite{Allanach:2002nj,AbdusSalam:2011fc}.
The scenario SPS1a has already been excluded by searches at LHC. However, it still 
serves as viable benchmark scenarios, where many detailed studies are available in the literature. The scenario 
10.1.6 is still viable and can be tested in the near future. The low energy spectrum for both scenarios has been obtained with the 
program \softsusy \cite{Allanach:2001kg}. Sparticle on-shell masses relevant for our analysis are listed in \tabref{tab:lowmasses}. Non-vanishing Yukawa corrections implemented in \softsusy result in small mass splittings between first and second generation squarks and sleptons. We verified that, for the CMSSM scenarios SPS1a and 10.1.6, the phenomenological effects originating from these small mass splittings are negligible in the study presented here. Thus, we set all second-generation masses equal to their first-generation counterparts.

\TABULAR[ht]{c||c|c|c|c|c|c|c|c|c}{
  \hline\trule
   & $\boldsymbol{\uL}$
  & $\boldsymbol{\uR}$
  & $\boldsymbol{\dL}$
  & $\boldsymbol{\dR}$
  & $\boldsymbol{\gluino}$
  & $\boldsymbol{\slep_L}$
  & $\boldsymbol{\slep_R}$
  & $\boldsymbol{\neu_2}$
  & $\boldsymbol{\neu_1}$
\\
  \hline
  SPS1a & $563.6$ & $546.7$ & $569.0$ & $546.6$ & $608.5$ & $202.4$ & $144.1$ & $180.2$ & $97.0$\\
   \hline
  10.1.6 & $1531.7$ & $1472.2$ &  $1533.6$ &  $1466.1$ & $1672.1$ & $536.6$& $340.6$ & $592.4$ & $313.3$ \\
  \hline
}{On-shell masses of the first generation squarks and sleptons, the gluino and the lightest and second lightest neutralino within the
different SUSY scenarios considered. All masses are given in \GeV. 
\label{tab:lowmasses}  
  }

In both scenarios the gluino is heavier then all light flavor squarks. Thus, all these squarks 
decay exclusively into charginos and neutralinos. In \tabref{tab:brachings} corresponding branching ratios, calculated with \sdecay \cite{Muhlleitner:2003vg}, are listed \footnote{In \tabref{tab:brachings} we list the average of the value of the branching ratios for up and down type squarks, that, however, differ at most by $\sim1\%$. Differences between branching ratios at LO and NLO for squark decays are negligible (less than per mill) for the considered scenarios and so not shown.}.
The right-handed squarks decay dominantly directly into the bino-like \neuone,the left-handed squarks into the \neutwo~and the lighter 
chargino, $\tilde{\chi}^{\pm}_1$.

\TABULAR[ht]{c||c|c|c|c|c|c|c}{
  \hline\trule
      $\textbf{BR (\%)}$
   & $\boldsymbol{\tilde{q}_{R}\rightarrow \neu_{1}}$
  & $\boldsymbol{\tilde{q}_{R}\rightarrow \neu_{2}}$
  & $\boldsymbol{\tilde{q}_{L}\rightarrow \neu_{1}}$
  & $\boldsymbol{\tilde{q}_{L}\rightarrow \neu_{2}}$
  & $\boldsymbol{ \neu_{2}\rightarrow \tilde{l}_{L}^{\pm}}$
  & $\boldsymbol{\neu_{2}\rightarrow \tilde{l}_{R}^{\pm}}$
    & $\boldsymbol{\neu_{2}\rightarrow  Z}$
\\
  \hline
  SPS1a & $ 98.5 $ & $1.0$ & $ 1.5 $ & $31.2$ & $ - $ & $13.1$ & - \\
   
   \hline
  10.1.6  & $99.8$ & $ 0.03 $ &  $1.5$ &  $32.1$ & $28.4$ & $0.2$ & $0.2$\\
   
  \hline
}{Branching ratios for the decay of squarks into $\neu_{1}$ and $\neu_{2}$ and for the decay of a $\neu_{2}$ into right- and left-handed 
sleptons. Squarks and leptons of the first two families are considered, where branchings into second and first generation sleptons and their 
charge-conjugate contributions are summed. 
\label{tab:brachings}  
  }

In \tabref{tab:brachings} we also list branching ratios for the second lightest neutralino into
light flavor sleptons. Branching ratios into first and second-generation sleptons are identical, 
and in \tabref{tab:brachings} we sum those contributions.
For benchmark point SPS1a only the right-handed $\slep_R$ is lighter then the \neutwo. Thus, next to the decay into
a $\tau$-slepton, this is the only available two-body decay. In our numerical analysis of SPS1a, both for the 
decay chain alone and combined with the production, only the decay via a right-handed 
slepton is considered. In contrast, for 10.1.6 both sleptons are lighter than the \neutwo. Due 
to its wino-like nature the \neutwo here decays dominantly into the left-handed $\slep_L$ despite the 
smaller mass of the $\slep_R$. For simplification in our numerical analysis of 10.1.6 only the decay via a left-handed 
slepton is considered. As can be seen from \tabref{tab:brachings}, the contribution from the decay into a $\slep_R$ (and also into $\neu_1 Z$) can be neglected safely.

In all numerical results presented in the following we employ the anti-$k_T$ jet clustering algorithm with a jet radius of $R=0.4$ implemented in \fastjet \cite{Cacciari:2011ma}. Furthermore we define a jet to fulfil the cut conditions
\begin{align}\label{eq:cuts_jets}
 p^T_{j_i} \ge 20~\GeV\, , \hspace{1cm} \vert \eta_{j_i} \vert \le 2.8 \, . 
\end{align}
Thus, we arrive at an experimentally well defined result. 
When analyzing combined production and decay in section \ref{sec:num_comb},   
the following realistic experimental cuts are applied:
\begin{align}\label{eq:cuts_combined}
 p^T_{j_1} & \ge 150~\GeV \,,& p^T_{j_2} &\ge 100~\GeV \,,& \vert \eta_{j,l} \vert \le 2.5 \,, \nonumber \\
 p^T_{l_{1,2}} & \ge 20~\GeV  ~\text{(OS-SF)}\,,& \missingET & \ge 100~\GeV \, ,
\end{align}
where we implicitly require the two leptons to have opposite charge and same flavor (OS-SF). Such cuts efficiently reduce SM backgrounds \cite{Weiglein:2004hn,Chatrchyan:2012te,Aad:2011xk}. Furthermore, we assume that contributions from leptonic decays
of $\tau$-leptons (from SM processes or the corresponding signal decay chain with an intermediate tau sleptons), charginos and $W^{\pm}$ bosons are removed in the standard way by subtracting events with opposite-sign different-flavor lepton pairs (OS-DF), see e.g. \cite{Hinchliffe:1996iu,Allanach:2000kt,Weiglein:2004hn,Gjelsten:2004ki,Barr:2010zj}. Thus these cuts help to isolate the decay chain under consideration.

\subsection{Squark decay chain}
\label{sec:num_chain}

\FIGURE{
\includegraphics[width=.49\textwidth]{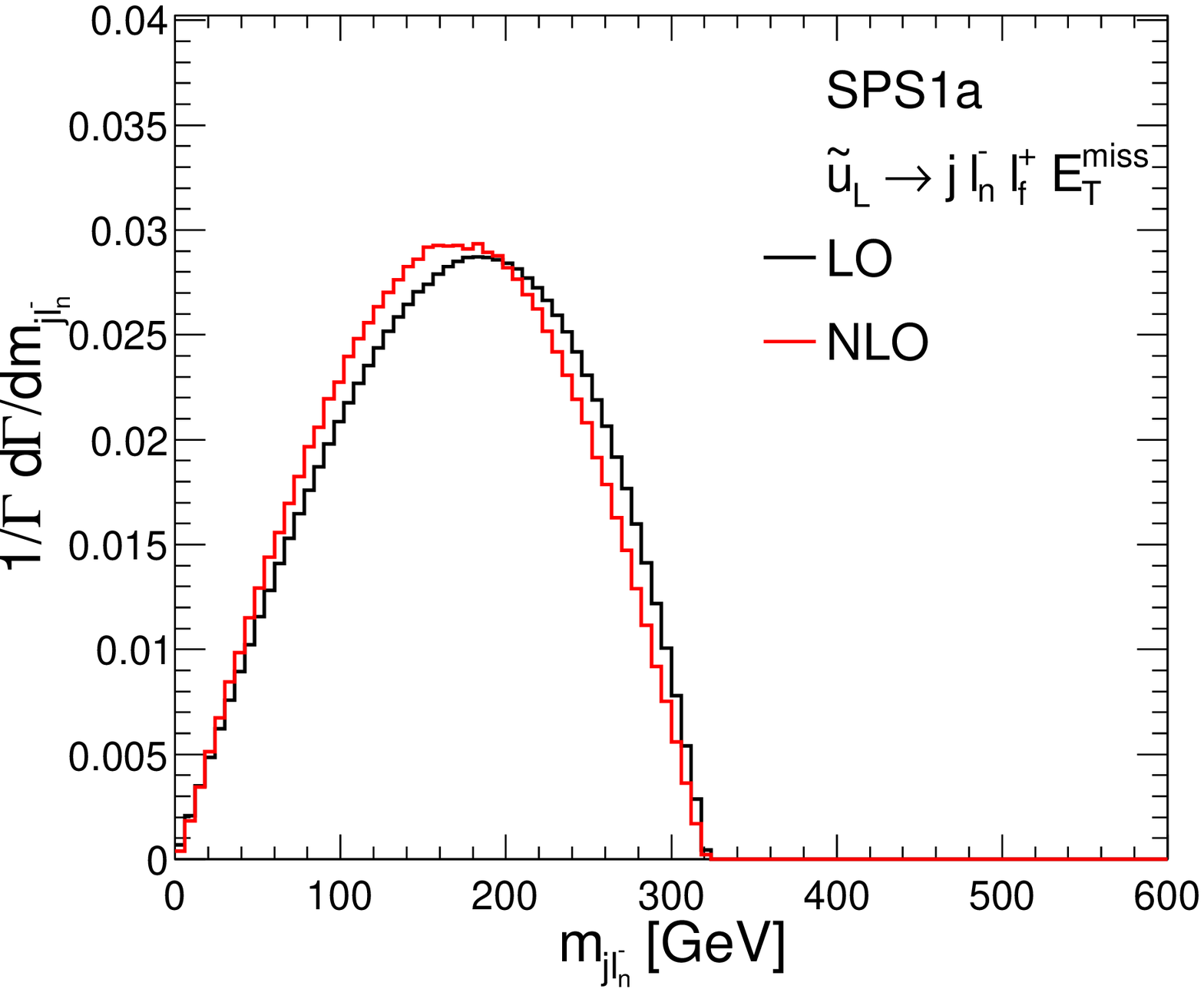}
\includegraphics[width=.49\textwidth]{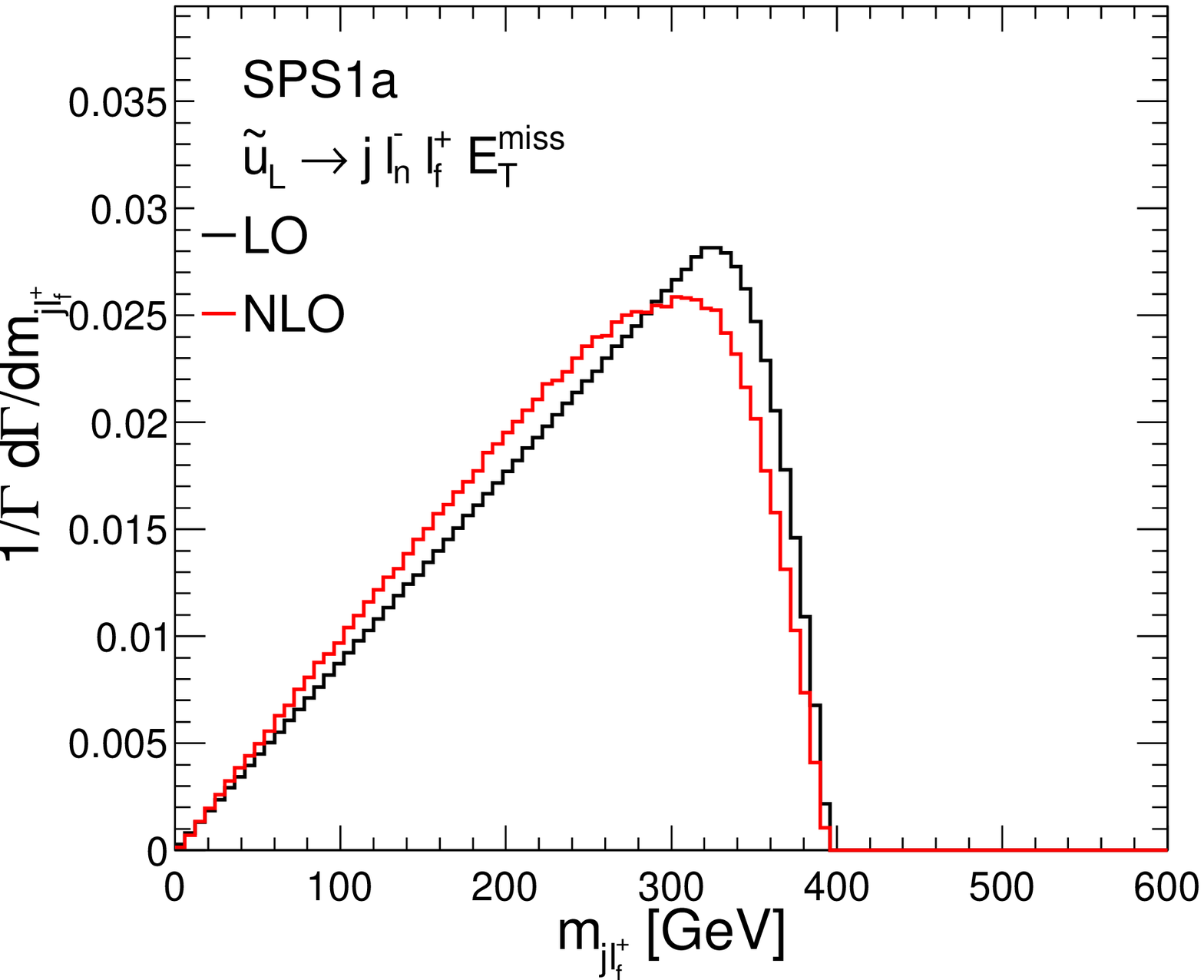}\\
\includegraphics[width=.49\textwidth]{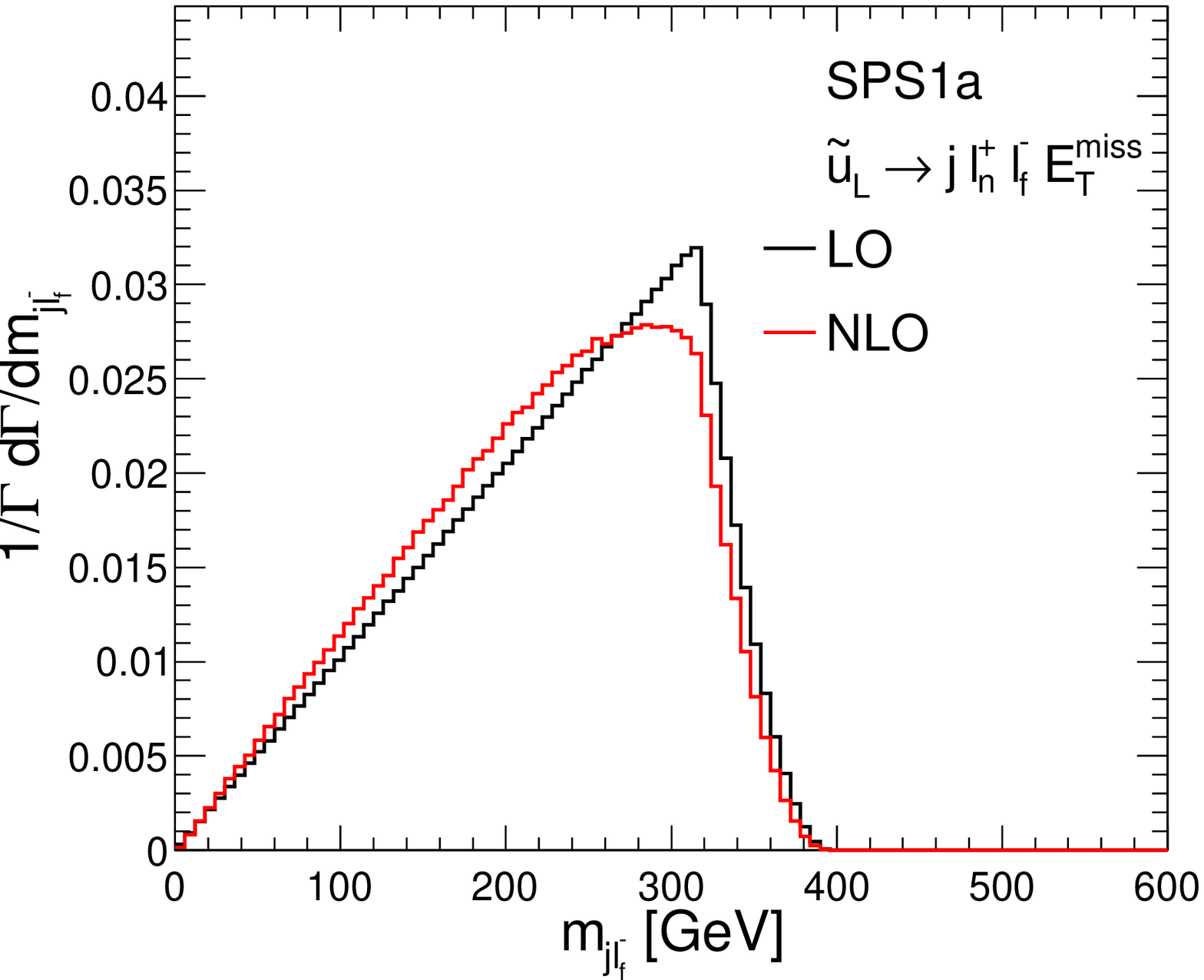}
\includegraphics[width=.49\textwidth]{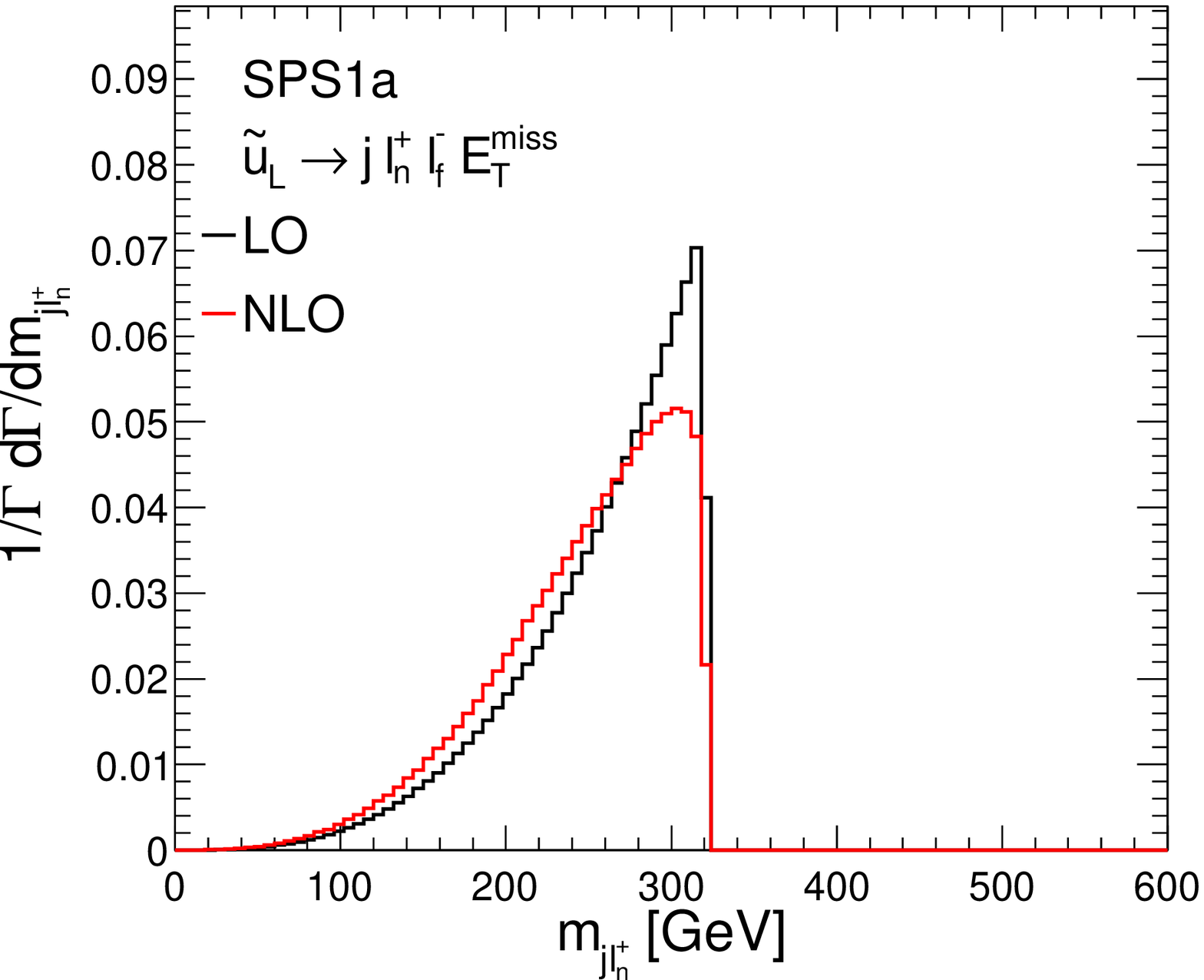}\\
\includegraphics[width=.49\textwidth]{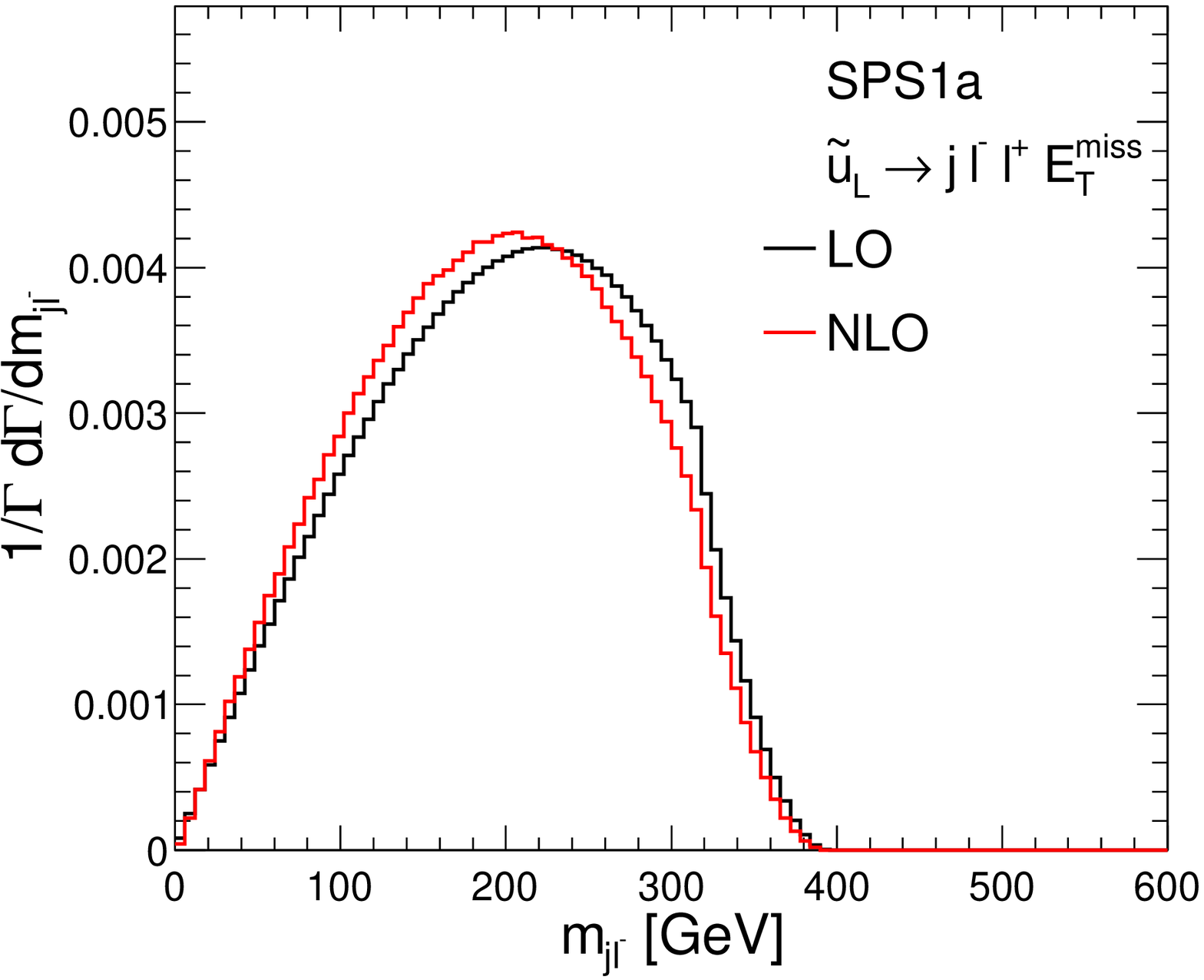}
\includegraphics[width=.49\textwidth]{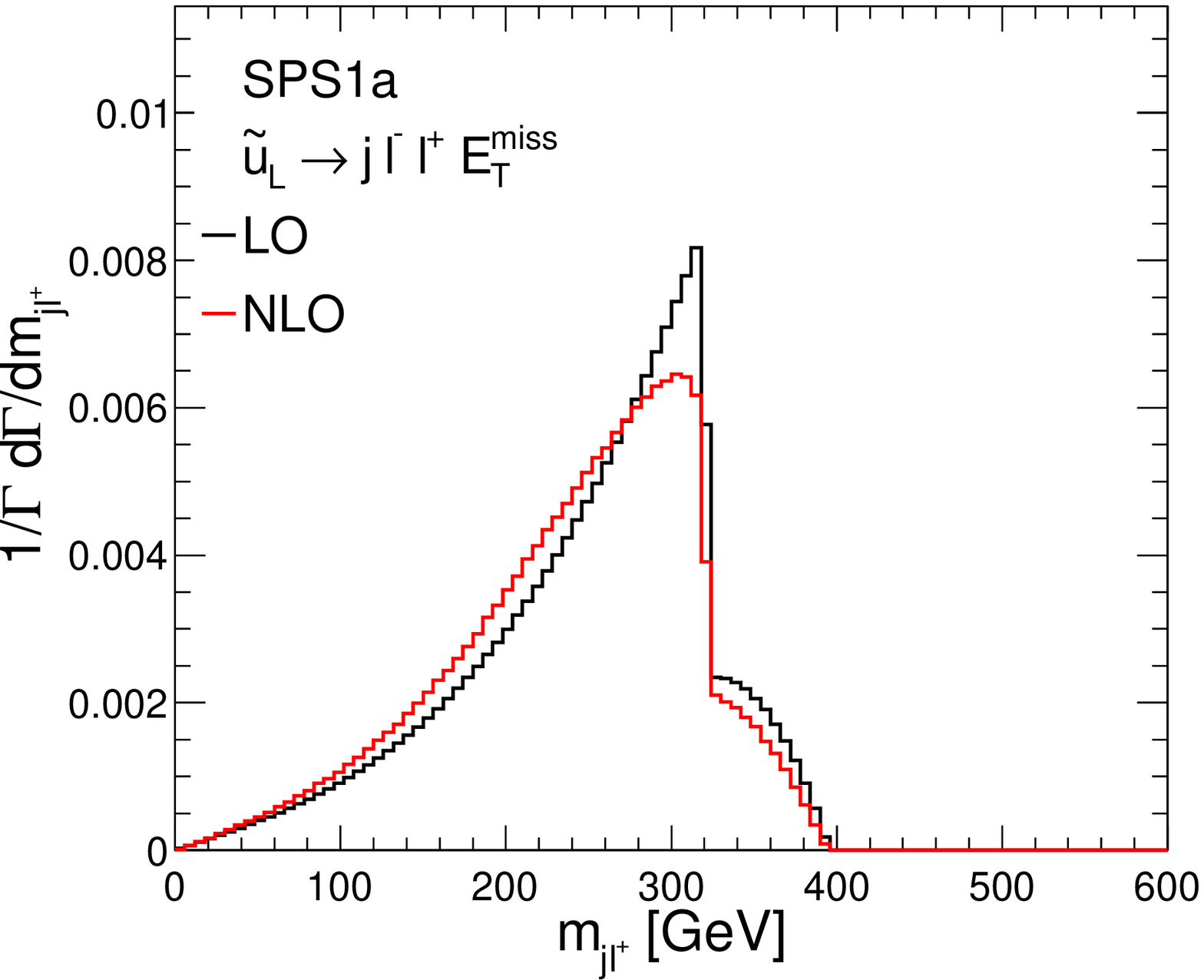}\\
\caption{Normalized differential distributions for SPS1a in $\mjlf$, $\mjln$ for the two unobservable decays $\tilde u_L \to j l_n^- l_f^+ \LSP$ (upper two) and $\tilde u_L \to j l_n^+ l_f^- \LSP$ (central two) and in $\mjlp$ and $\mjlm$ (lower two) where contributions from the two decays are summed. LO predictions are shown in black, NLO in red.
}
\label{fig:chain_sps1a}
}

%
Here we want to investigate NLO corrections to the isolated decay chain \eqref{qLchain} evaluated in the squark rest frame. 
In \figref{fig:chain_sps1a} we show various invariant mass distributions in the final state leptons and jet(s) for the benchmark scenario SPS1a, LO distributions are shown in black and NLO distributions in red. As explained in section \ref{sec:intro}, shapes of such distributions are very important for the determination of masses and spins of sparticles. In order to highlight NLO corrections purely in the shapes, here and in section \ref{sec:num_comb} we show all distributions, both at LO and at NLO, normalized to unity.

Two kinds of combinatorial problems arise in studying invariant mass distributions involving the final state leptons and jet(s). First, as already mentioned in section \ref{sec:intro}, from an experimental point of view we cannot distinguish between the near and the far lepton on an event-by-event basis. This is a well known problem and many solutions have been suggested in the literature, see e.g.~\cite{Barr:2010zj}. Second, it is not obvious which jet to choose to built the desired invariant mass distributions. Considering only the isolated decay chain starting with $\sq_L$, at LO only the jet from the squark decay is present in the final state. But at higher orders, due to real gluon radiation, further jets can be present. Here we always choose the hardest available jet to build the invariant mass distributions,
as done in~\cite{Horsky:2008yi}.

In the upper left/right part of \figref{fig:chain_sps1a} we show (unobservable) distributions in the invariant mass of the hardest jet and the negatively/positively charged near/far lepton from the decay chain $\tilde u_L \to j l_n^- l_f^+ \LSP$. Here and in the remainder of this section we do not apply any cuts, but the jet definition cuts in \eqref{eq:cuts_jets}. In the center left/right part, on the other hand, we show (again unobservable) distributions in the invariant mass of the hardest jet and the negatively/positively charged far/near lepton from the decay chain with the opposite charges for far and near leptons, $\tilde u_L \to j l_n^+ l_f^- \LSP$. 
Finally, in the lower part of \figref{fig:chain_sps1a} we show, in some sense, the sums of the two previous contributions. These distributions are in principle experimentally observable (after combination with the corresponding production process). The lower left/right panel shows the invariant mass of the hardest jet and the negatively/positively charged lepton summed over near and far contributions ($\mjlm$ and $\mjlp$). 

In the case of the decay of a left-handed anti-squark, all the distributions introduced so far are equal to the charge-conjugate ones of the corresponding squark, e.g, the $m_{jl^{+}_{n}}$ distribution from an $\tilde{u}^{*}_{L}$ decay chain is equal to the $m_{jl^{-}_{n}}$ distribution from an $\tilde{u}_{L}$ decay chain. Hence, the analogue of \figref{fig:chain_sps1a} for  $\sq_L^*$ would present the shapes of the distributions of the left column exchanged with the ones of the right column. These differences between squarks and anti-squarks obviously do not appear for quantities that are inclusive in the different charges of the leptons.

In all of the plots in \figref{fig:chain_sps1a} NLO corrections tend to shift the distributions to smaller invariant masses, however, locations of endpoints are unaffected. Kinematical edges in the NLO predictions are rounded off compared to LO, still, overall shapes of the considered contributions seem to be unaltered. Results and distributions of the same type, as already stated, have been calculated in \cite{Horsky:2008yi}. In their numerical evaluation the slightly different parameter point SPS1a' was investigated resulting 
in LO shapes somewhat different to those presented here. 
Qualitatively the NLO corrections shown in \cite{Horsky:2008yi} for SPS1a' and ours for  SPS1a agree.
Moreover, we also investigated distributions and corrections for SPS1a' and found  agreement with~\cite{Horsky:2008yi}\footnote{In reference \cite{Horsky:2008yi} a different jet algorithm is used. Results for SPS1a' presented there for $y_c=0.002$ agree best with our results obtained using the anti-$k_T$ jet clustering algorithm.}.\\

\FIGURE[t]{
\includegraphics[width=.49\textwidth]{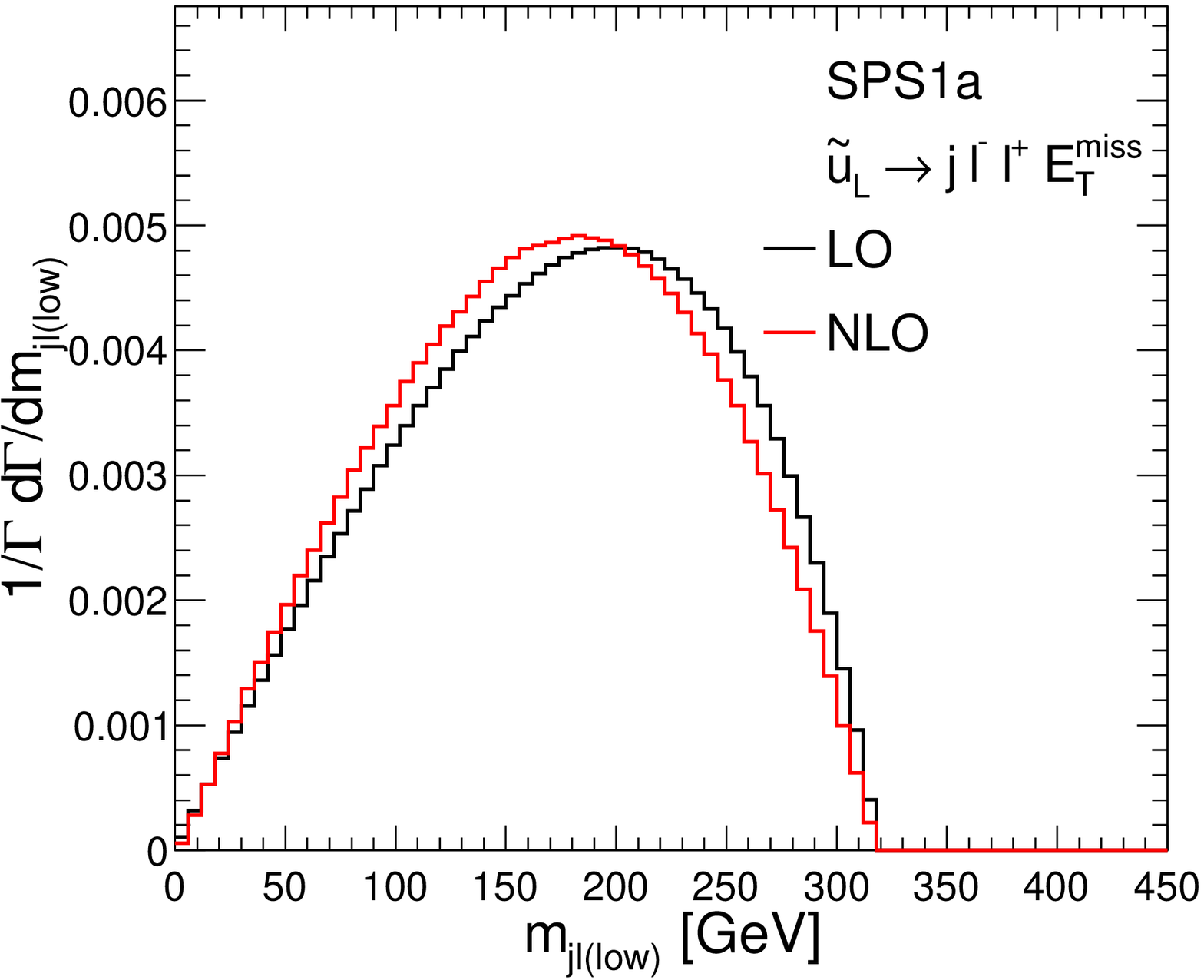}
\includegraphics[width=.49\textwidth]{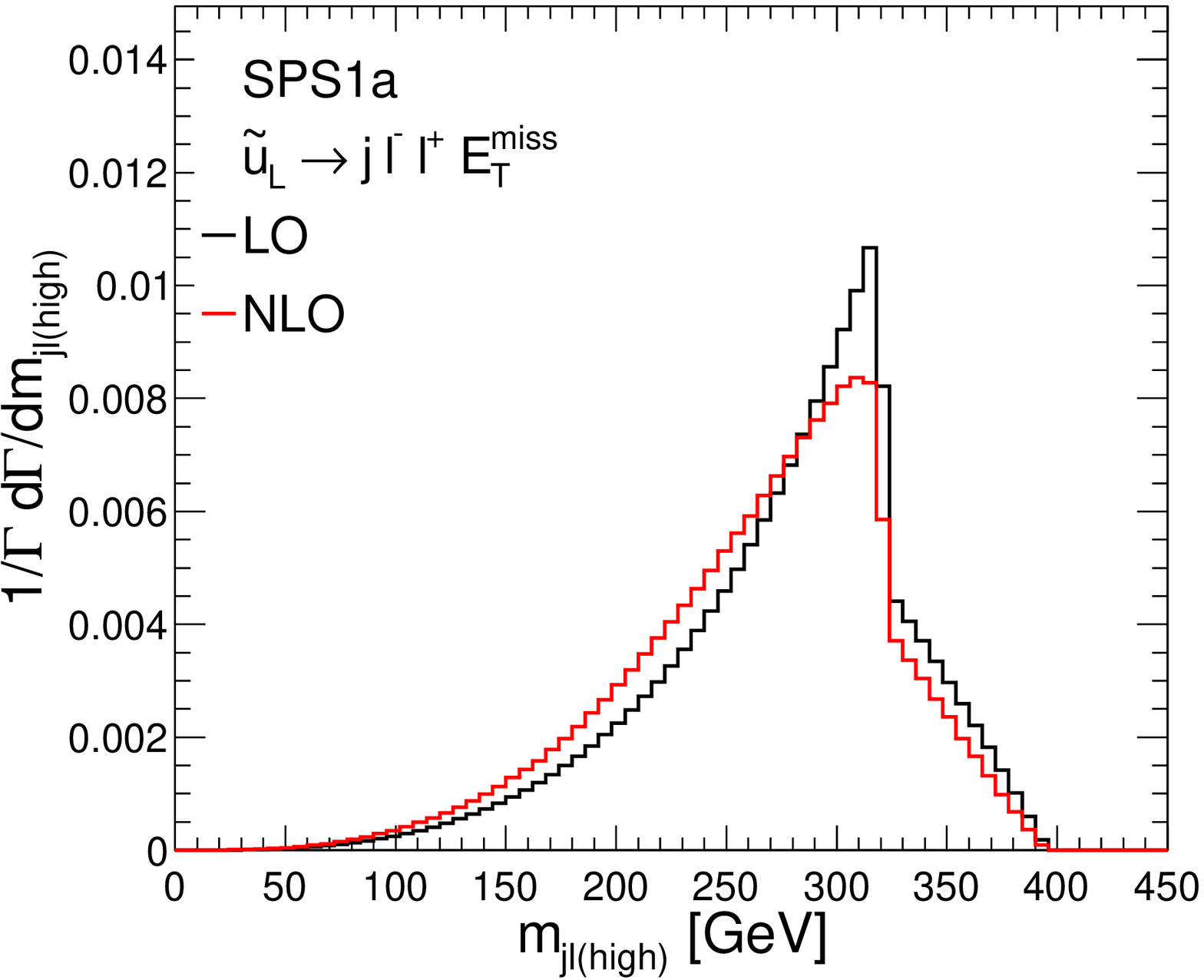}\\
\includegraphics[width=.49\textwidth]{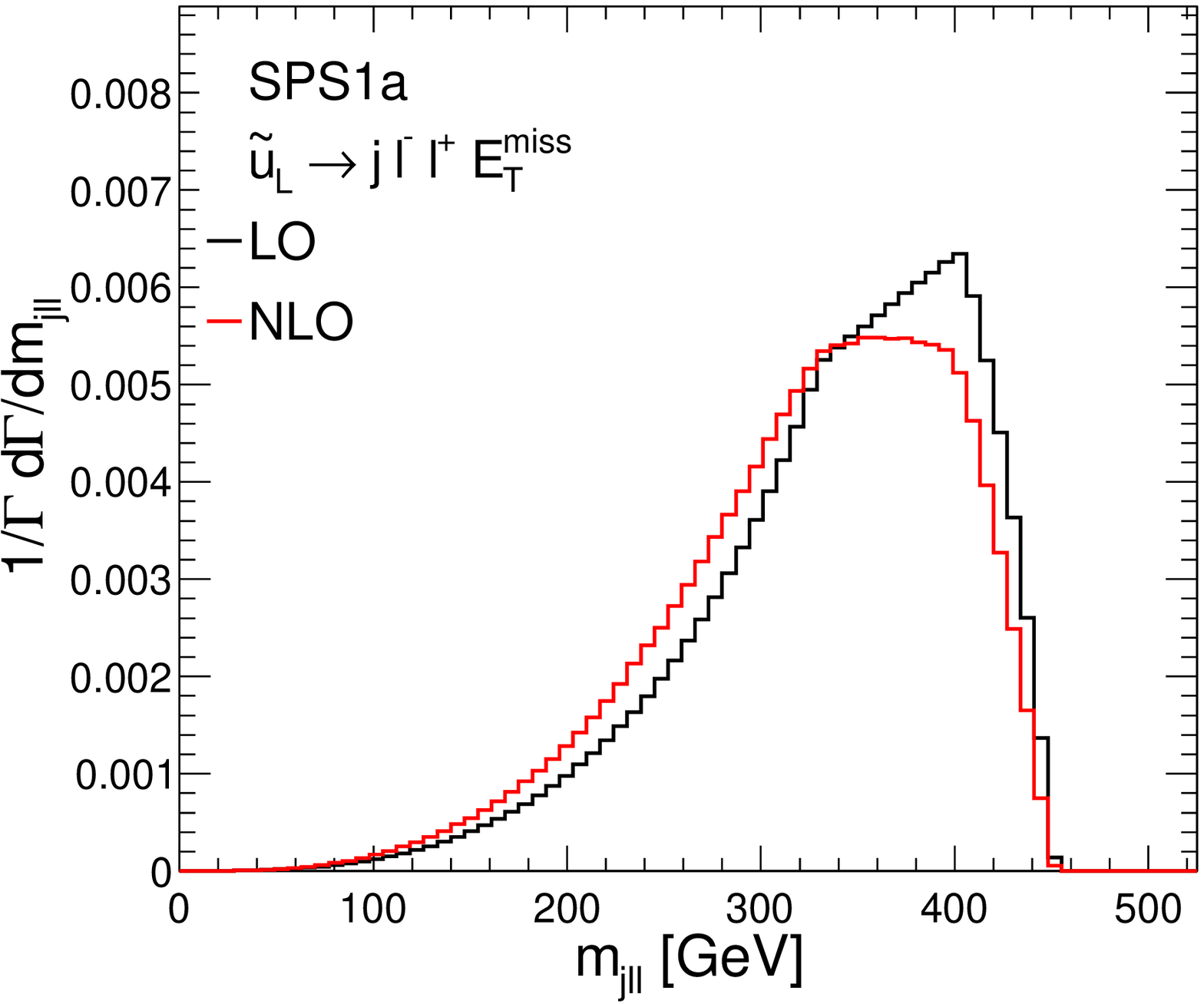}
\includegraphics[width=.49\textwidth]{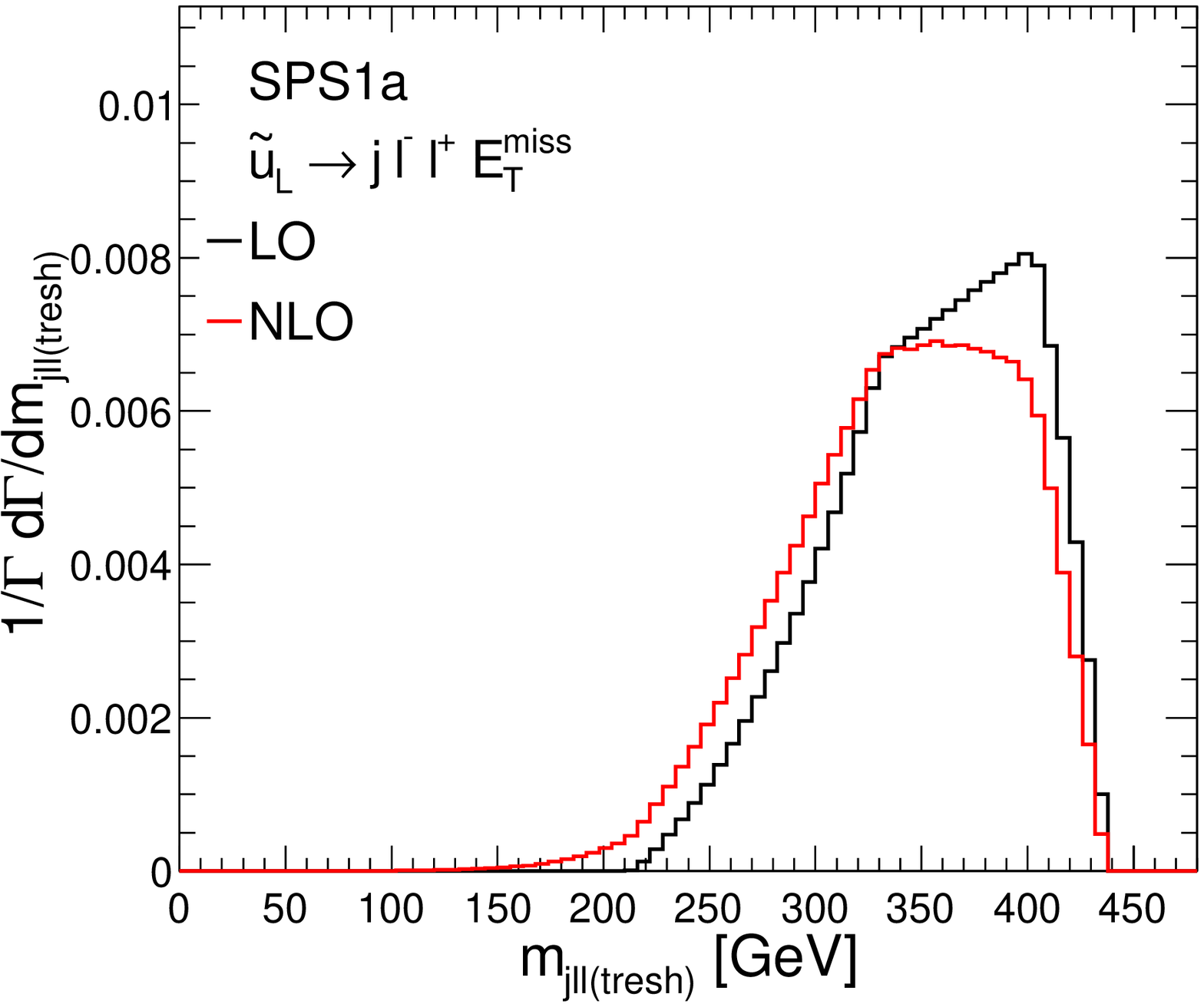}\\
\caption{LO (black) and NLO (red) normalized differential distributions for SPS1a in $\mjllow$, $\mjlhigh$, $\mjll$ and $\mjlltresh$ (from top left to bottom right) for the decay chain $\qLRchain$. 
}
\label{fig:chain_sps1a_2}
}

In \figref{fig:chain_sps1a_2} we look at a different set of invariant mass distributions again for SPS1a: $\mjllow$, $\mjlhigh$, $\mjll$ and $\mjlltresh$. For $\mjllow$ and $\mjlhigh$ we select on an event-by-event basis the smaller and higher invariant mass between one of the leptons and the hardest jet. $\mjll$ is the invariant mass between the hardest jet and the two leptons and $\mjlltresh$ is the same distribution where an additional constraint on the invariant mass of the two leptons, $\frac{m_{ll}^{\text{max}}}{\sqrt{2}} < m_{ll}$, is applied. Here, $m_{ll}^{\text{max}}$ is the well measurable endpoint of the dilepton invariant mass distribution \footnote{In our numerical analysis we use the theoretical endpoints $m_{ll}^{\text{max}}=80.0,203.8~\GeV$ for SPS1a and 10.1.6.}. All these invariant mass distributions are in principle experimentally measurable and have extensively been discussed in the literature \cite{Hinchliffe:1996iu,Weiglein:2004hn,Gjelsten:2004ki,Barr:2010zj} (and references therein). 
From a measurement of their upper (and in case of $\mjlltresh$ the lower) endpoints one might be able to extract relations for the masses of the intermediate sparticles. These relations often show ambiguities and particularly measurements like the threshold of $\mjlltresh$ might help to resolve these. Alternatively, shapes of the presented invariant mass distributions might help to overcome these difficulties. Let us now look at the NLO corrections to these invariant mass distributions. Overall, again, distributions are shifted to smaller  invariant masses. Also, upper kinks of $\mjll$ and $\mjlltresh$ are rounded off. These shifts might result in a slightly lowered accuracy in the measurement of the upper endpoints. Furthermore, the threshold of the $\mjlltresh$ is diluted due to NLO corrections\footnote{
The theoretical lower endpoint of the $\mjlltresh$ distribution is given by $\mjlltresh^{\text{min}}=215.4$ for SPS1a and $\mjlltresh^{\text{min}}=437.1$ for 10.1.6 \cite{Allanach:2000kt}.}. A precise measurement of this observable seems to be questionable.

The same set of invariant mass distributions, $\mjllow$, $\mjlhigh$, $\mjll$ and $\mjlltresh$, is shown in \figref{fig:chain_1016_2} for the parameter point 10.1.6 and the corresponding decay chain involving a $\slep_L$. For the main part, again, NLO corrections shift the differential distributions to smaller invariant masses and round off the upper kinks. Particularly for $\mjllow$ this might result in a smaller possible experimental accuracy for determining the upper endpoint. Apart from the rounding off of the kinks, general shapes of the distributions are mostly unaltered by NLO corrections. Also, a possible dilution of the lower endpoint of $\mjlltresh$ due to NLO corrections seems to be less severe for 10.1.6 compared to SPS1a.

\FIGURE[t]{
\includegraphics[width=.45\textwidth]{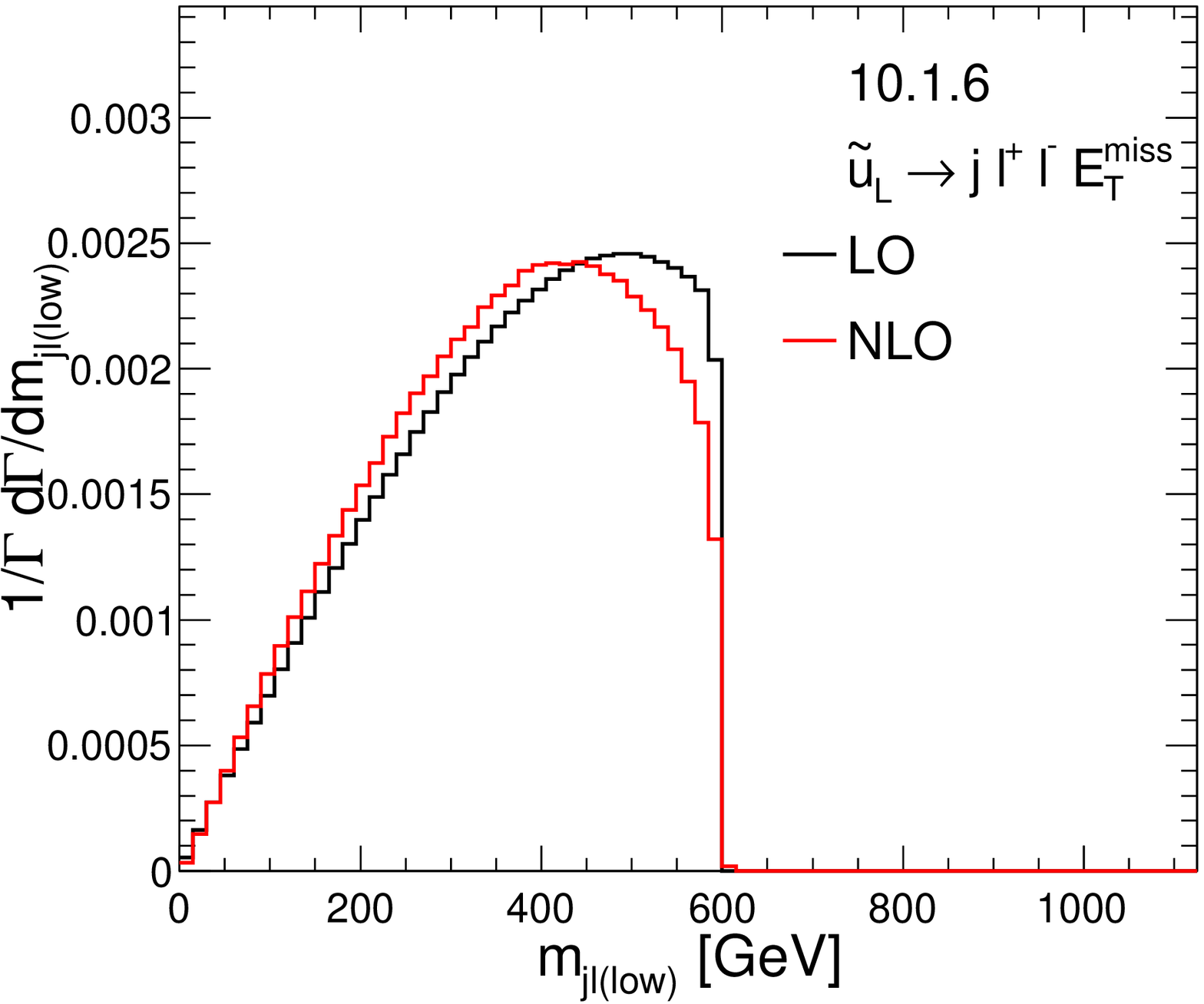}
\includegraphics[width=.45\textwidth]{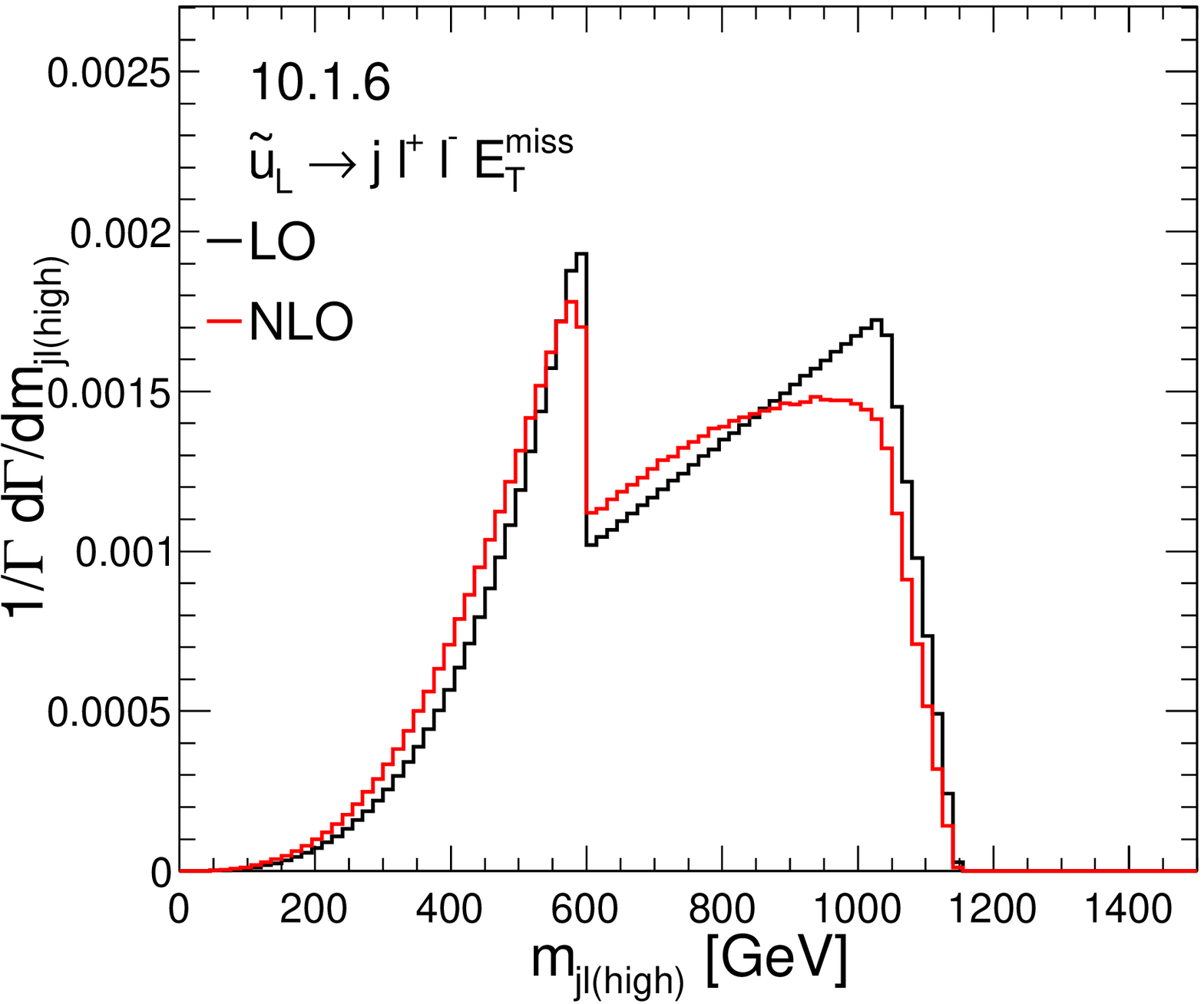}\\
\includegraphics[width=.45\textwidth]{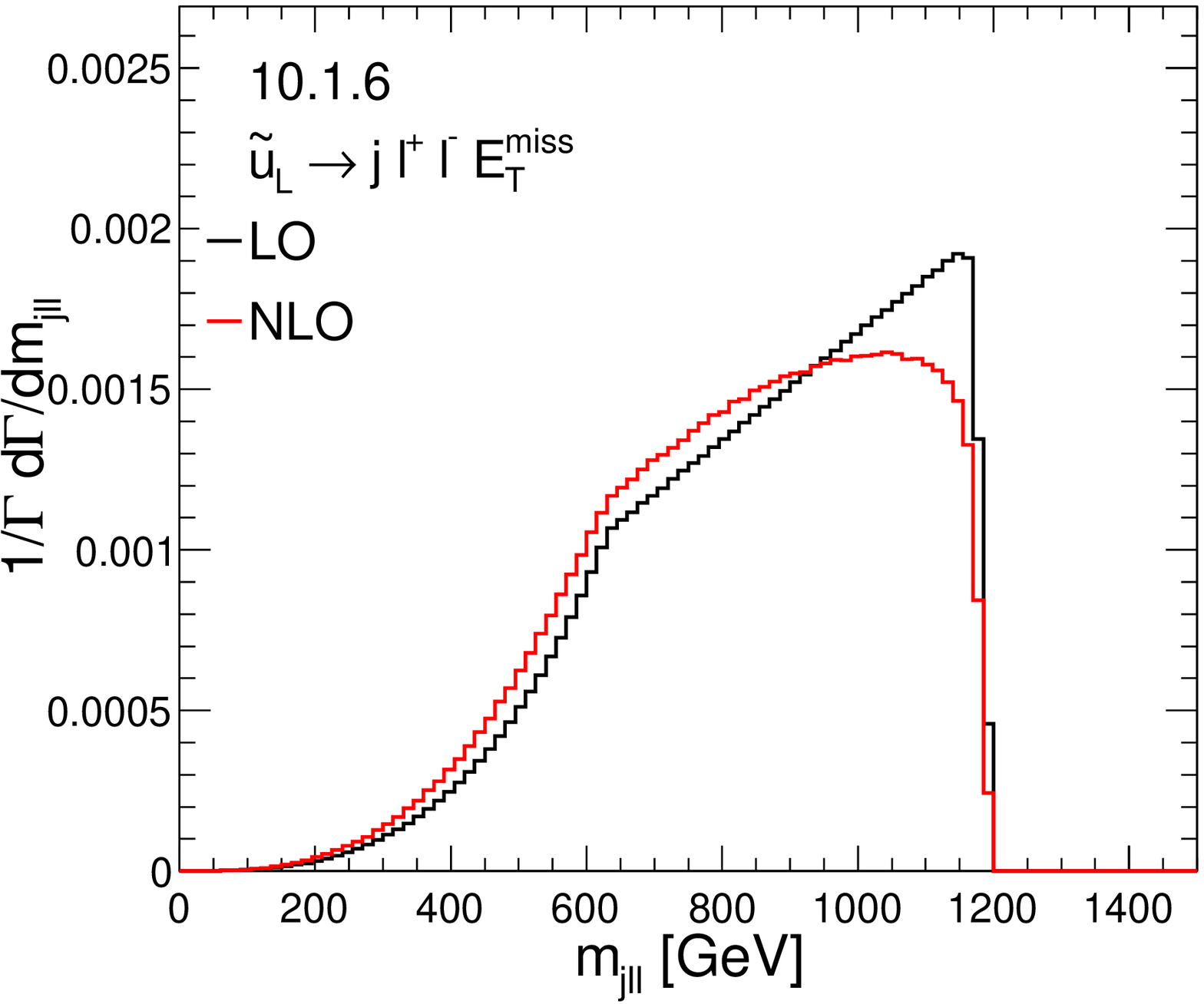}
\includegraphics[width=.45\textwidth]{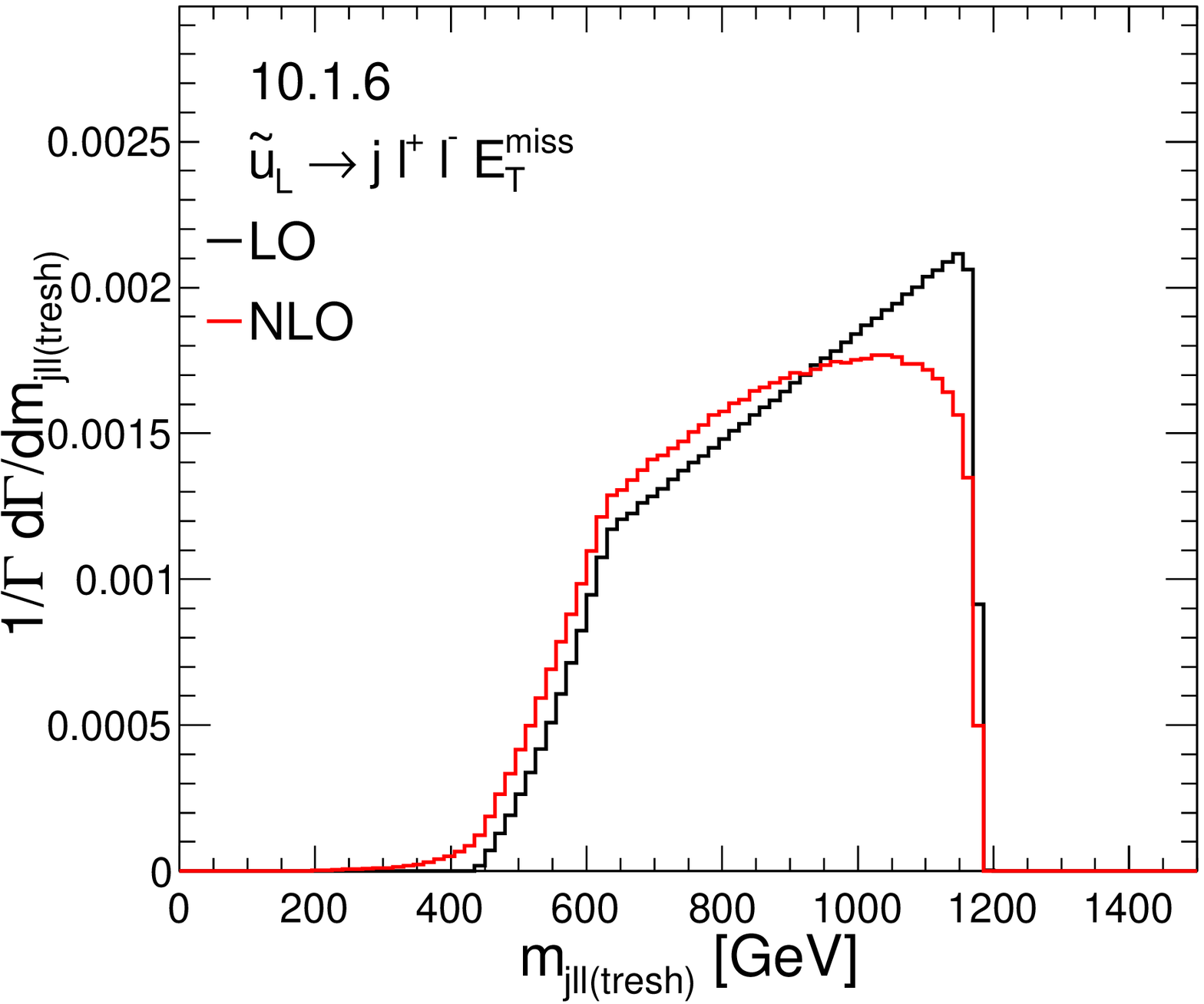}\\
\caption{LO (black) and NLO (red) normalized differential distributions for 10.1.6 in $\mjllow$, $\mjlhigh$, $\mjll$ and $\mjlltresh$ (from top left to bottom right) for the decay chain $\qLLchain$. 
}
\label{fig:chain_1016_2}
}

\pagebreak

\subsection{Combination of production and decay}
\label{sec:num_comb}
Now we want to investigate the combined process, \eqref{eq:process}, where cuts defined in \eqref{eq:cuts_combined} have been applied. In section \ref{sec:num_dist} we first examine the impact on various differential distributions important for parameter determination. Afterwards, in section \ref{sec:num_inc} we investigate the impact of NLO QCD corrections on inclusive OS-SF dilepton observables and thus on searches currently performed at the LHC.   

\subsubsection{Invariant mass distributions}
\label{sec:num_dist}
When combining production and decay already at LO a combinatorial problem arises when looking at invariant mass distributions. As already mentioned in section \ref{sec:num_chain} it is not clear which jet to choose for building the different invariant masses. Just choosing the hardest jet as was done in section \ref{sec:num_chain} does not seem to be sensible, since the jet from the second decay, present already at LO, is often the hardest one.
This is a well known problem in the application of the endpoint methods for mass determination and various methods have been developed to reduce this ambiguity. The easiest method is to always use the jet which gives e.g. the smaller $\mjll$ value. In this way one improves the measurements of the upper endpoints without losing statistics, however, shapes are heavily distorted already at LO.
In principle there are advanced techniques to solve this problem, amongst others, full kinematic event reconstruction \cite{Cheng:2007xv} or hemisphere techniques \cite{Baer:1995nq,Ball:2007zza,Nojiri:2008hy}. 
However, these techniques are quite involved, parameter point dependent and 
not generically applicable. Here, we apply consistency cuts, also discussed in \cite{Gjelsten:2004ki} to reduce the impact of the jet combinatorics ambiguity. 
Applying such consistency cuts means, we accept only events where one jet $j_i$ out of the two hardest jets $j_{i},j_{k}$ gives an invariant mass smaller than $m_{jll}^{\text{max}}$ and the other jet $j_k$ an invariant mass larger than $m_{jll}^{\text{max}}$,
\begin{align}\label{consistency}
 m_{j_i ll} < m_{jll}^{\text{max}} < m_{j_kll}~.
\end{align}
Now $j_i$ will be used in the following to build the invariant mass distributions. 
This technique%
\footnote{From an experimental point of view the endpoint $m_{jll}^{\text{max}}$ is assumed to be measured in a first step where for example always the jet is chosen yielding the smaller \mjll. Here, we use the theoretical endpoints $\mjll^{\text{max}}=450.6,1147.7~\GeV$ for SPS1a and 10.1.6.}
is very efficient in reducing the jet combinatorics ambiguity, however, event rates are also reduced (see section \ref{sec:num_inc} and particularly \tabref{tab:cmsrates}).\\

\FIGURE{
\includegraphics[width=.45\textwidth]{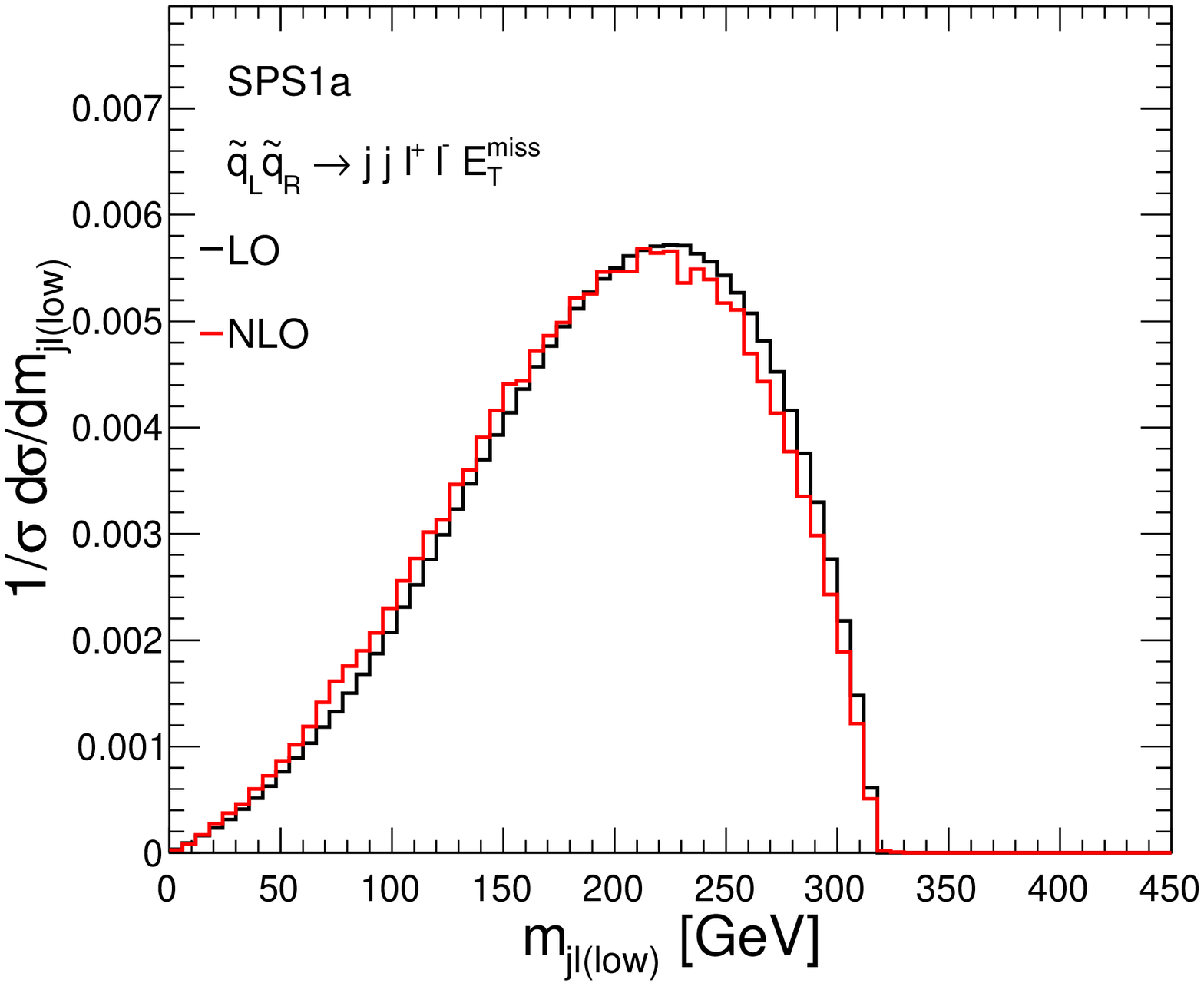}
\includegraphics[width=.45\textwidth]{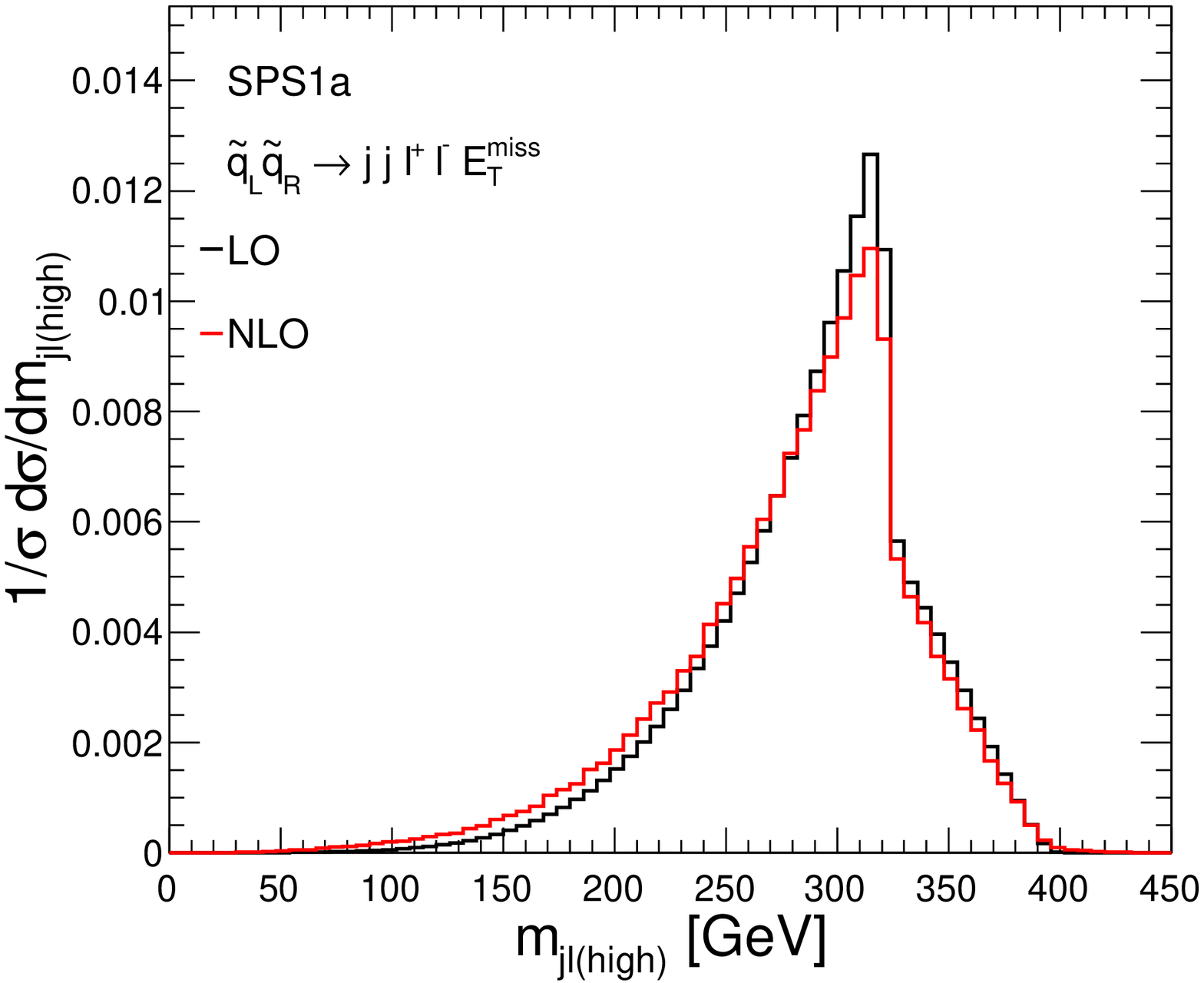}\\
\includegraphics[width=.45\textwidth]{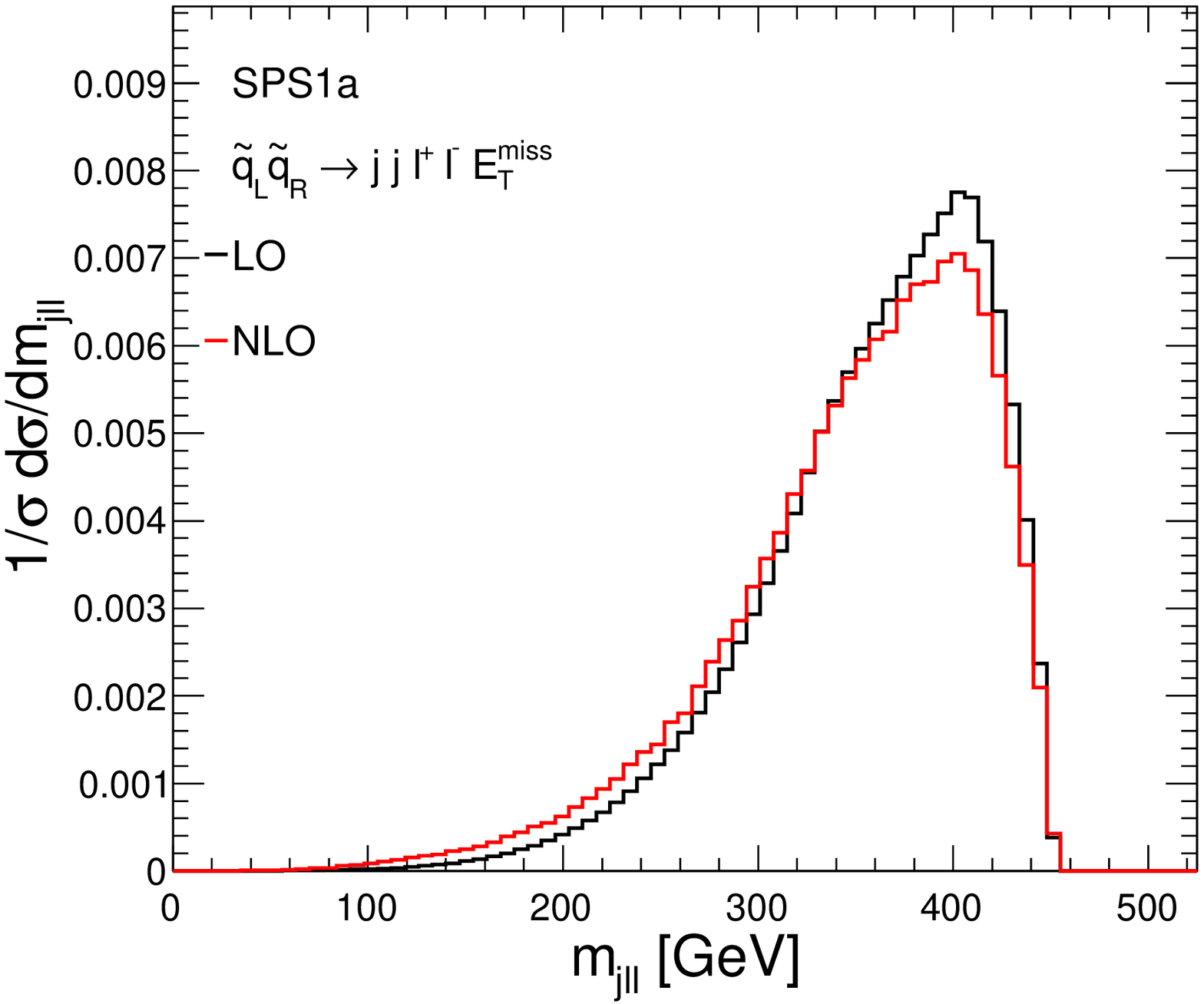}
\includegraphics[width=.45\textwidth]{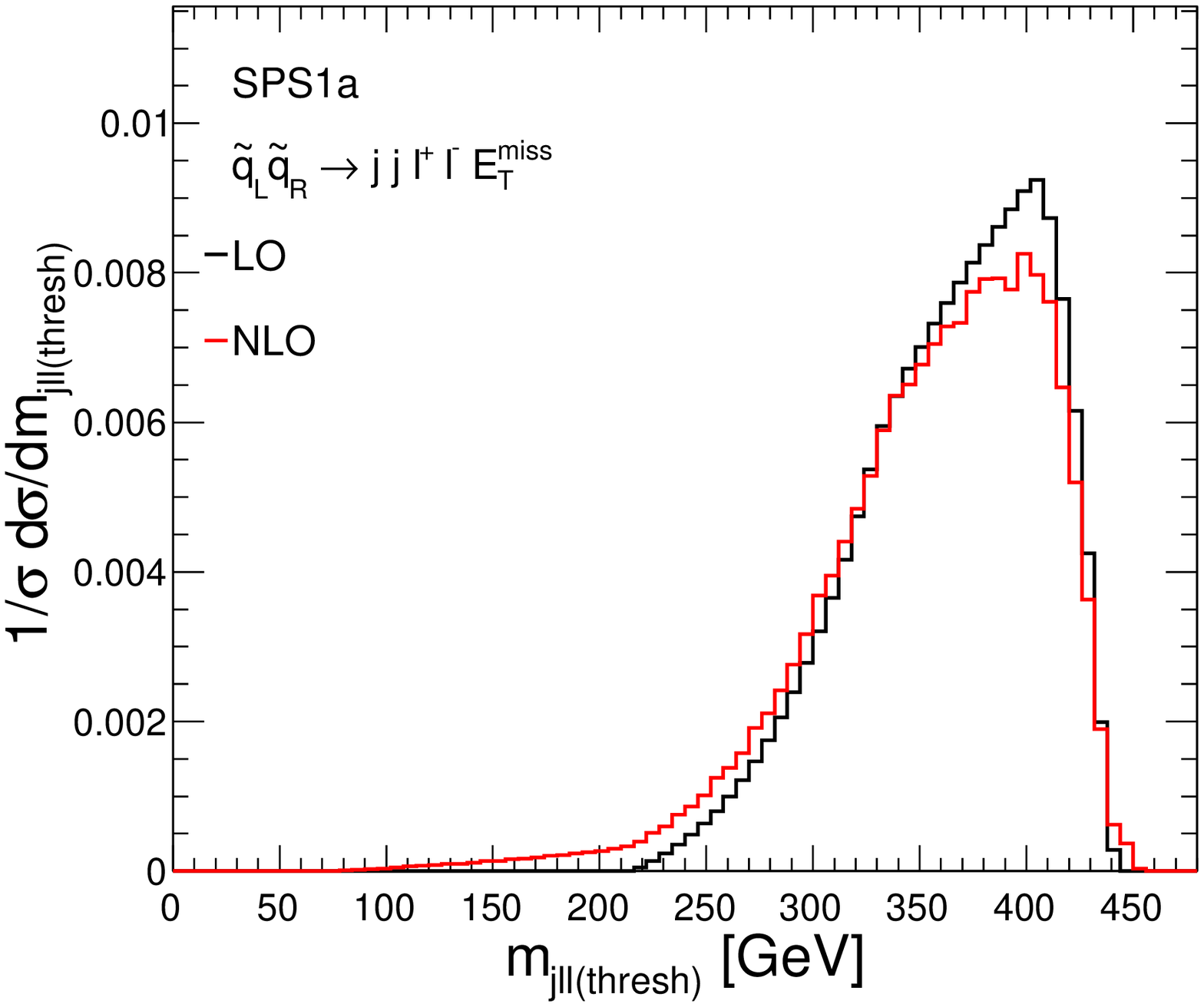}\\
\caption{LO (black) and NLO (red) normalized differential distributions for SPS1a in $\mjllow$, $\mjlhigh$, $\mjll$ and $\mjlltresh$ (from top left to bottom right) for combined production and decay.
}
\label{fig:combined_sps1a}
}

In \figref{fig:combined_sps1a} we show for SPS1a the same invariant mass distributions as already shown in \figref{fig:chain_sps1a_2}. Here, production and decays are combined at NLO, cuts of \eqref{eq:cuts_combined} and further consistency cuts are applied. 
Due to the NLO corrections distributions are in general shifted to smaller invariant masses. Comparing just LO predictions in \figref{fig:combined_sps1a} and \figref{fig:chain_sps1a_2} particularly $\mjll$ and $\mjlltresh$ show a slightly different behavior introduced by the consistency cuts: the plateau is less prominent in \figref{fig:combined_sps1a}. Here, again we observe a dilution of the threshold in the $\mjlltresh$ distribution at NLO. Similar observations can be made looking at \figref{fig:combined_1016} (and  comparing with \figref{fig:chain_1016_2}) for the combined results of parameter point 10.1.6. Overall, changes to the shapes are moderate and, concerning the measurements of the upper endpoints only for $\mjllow$ an 
experimentally detectable effect is expected. Here we want to note one thing: the consistency cuts, \eqref{consistency}, are based only on $\mjll$ and this is why, already at LO, we observe contributions also beyond the theoretical upper endpoint. This effect is enhanced at NLO.

From the discussion above usefulness of the threshold of $\mjlltresh$ seems questionable. Additionally, a measurement of a lower endpoint is always subject to large experimental backgrounds \cite{Lester:2006yw,Matchev:2009iw}. As this threshold was introduced to solve ambiguities in the mass determination due to the near-far indistinguishability, new techniques for this purpose have been invented.
In \cite{Matchev:2009iw} the authors argue that all invariant mass distributions used for mass determination from the given decay chain should be built symmetrically under the interchange $l^{\text{near}} \leftrightarrow l^{\text{far}}$. In this spirit they introduce a new set of invariant mass distributions,
$m^2_{jl(u)} \equiv m^2_{jl_n} \cup m^2_{jl_f}$,
$m^2_{jl(d)} \equiv | m^{2}_{jl_n}-m^{2}_{jl_f}|$,
$m^2_{jl(s)} \equiv m^{2}_{jl_n}+m^{2}_{jl_f}$ and
$m^2_{jl(p)} \equiv m_{jl_n} \cdot m_{jl_f}. $ 
Here we study the impact of the NLO QCD corrections on this class of distributions.
In \figref{fig:combined_1016_matchev} we show the normalized LO and NLO distributions in 
$m^2_{jl(u)}, m^2_{jl(d)}, m^2_{jl(s)}$ and $m^2_{jl(p)}$
against a quadratic scale. Shapes of these distributions are slightly changed due to NLO corrections, however, the possibility of measuring their upper endpoints (both endpoints in the case of $\mjlu$) seems to be unaffected. \\

Besides for mass determination, the given decay chain can also be used for spin determination or, more precisely, for spin distinction. As pointed out in \cite{Barr:2004ze} and many subsequent works, the asymmetry between the $\mjlp$ and $\mjlm$ distributions  defined as
\begin{align}
 A = \frac{d\sigma/d\mjlp - d\sigma/d\mjlm}{d\sigma/d\mjlp + d\sigma/d\mjlm}
\end{align}
can help to discriminate between a SUSY model and other models like Universal Extra Dimensions (UED) with a similar decay chain where the intermediate particles have different spins \cite{Smillie:2005ar}. In \figref{combined_A} we show LO (black) and NLO (red) predictions for this asymmetry for SPS1a (left) and 10.1.6. Again, cuts of \eqref{eq:cuts_combined} and consistency cuts have been applied. The potential for spin determination and/or model discrimination seems to be unaltered by NLO QCD corrections. For SPS1a at NLO there is a contribution beyond the upper endpoint not present for this observable at LO. This NLO contribution passes the consistency cuts which are based on the endpoint of the $m_{jll}$ distribution. However, expected event rates in this region of the asymmetry distribution are experimentally negligible.

\FIGURE{
\includegraphics[width=.45\textwidth]{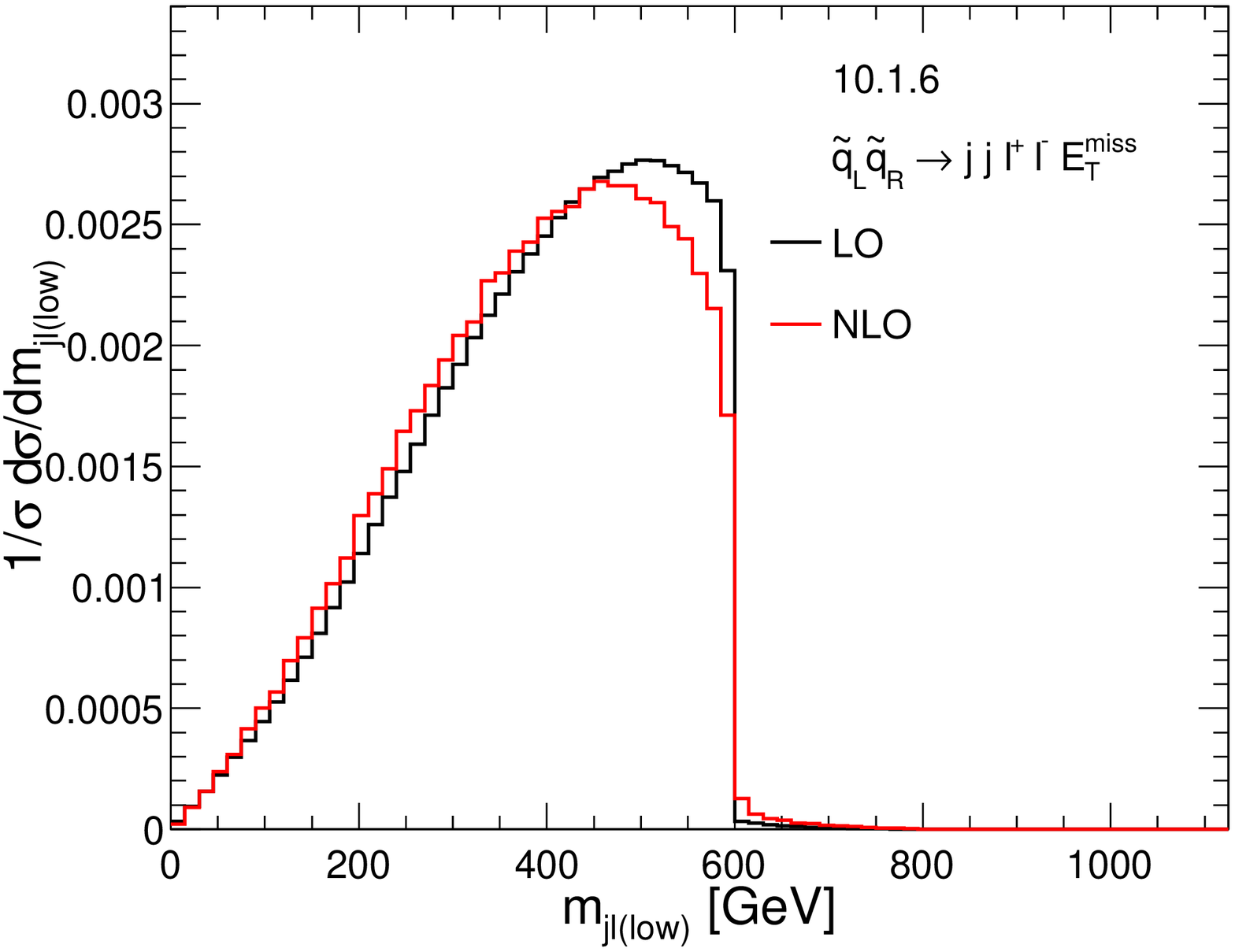}
\includegraphics[width=.45\textwidth]{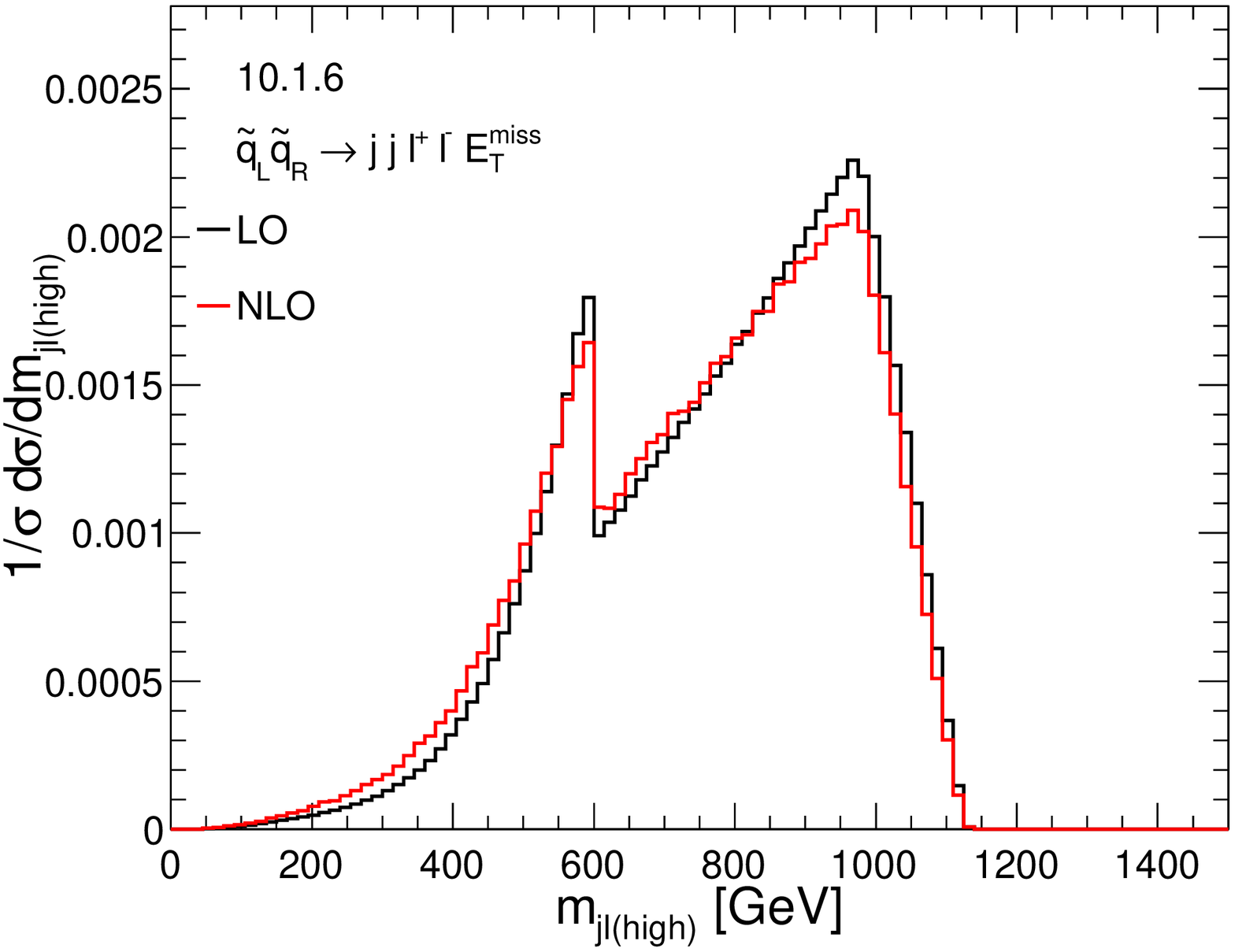}\\
\includegraphics[width=.45\textwidth]{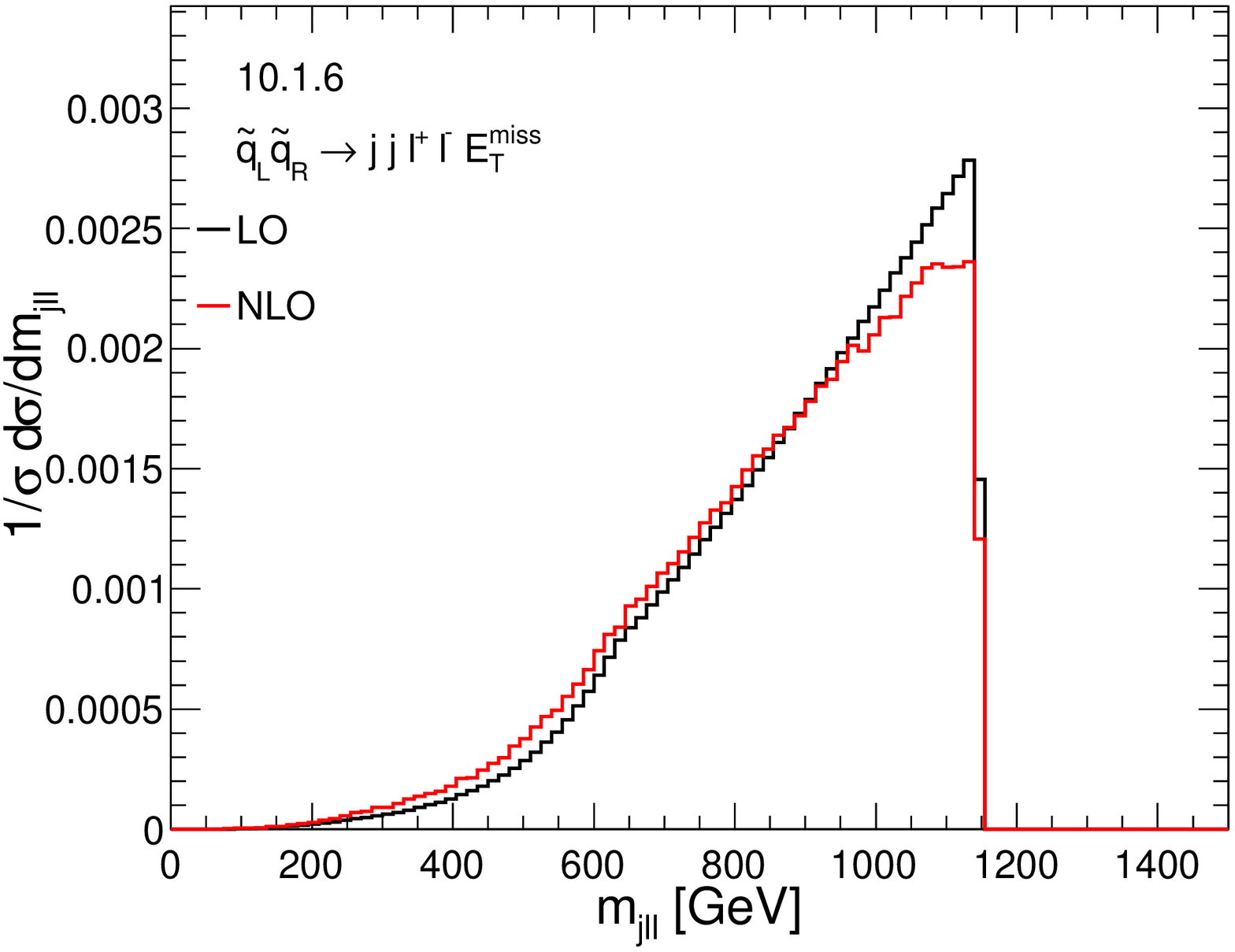}
\includegraphics[width=.45\textwidth]{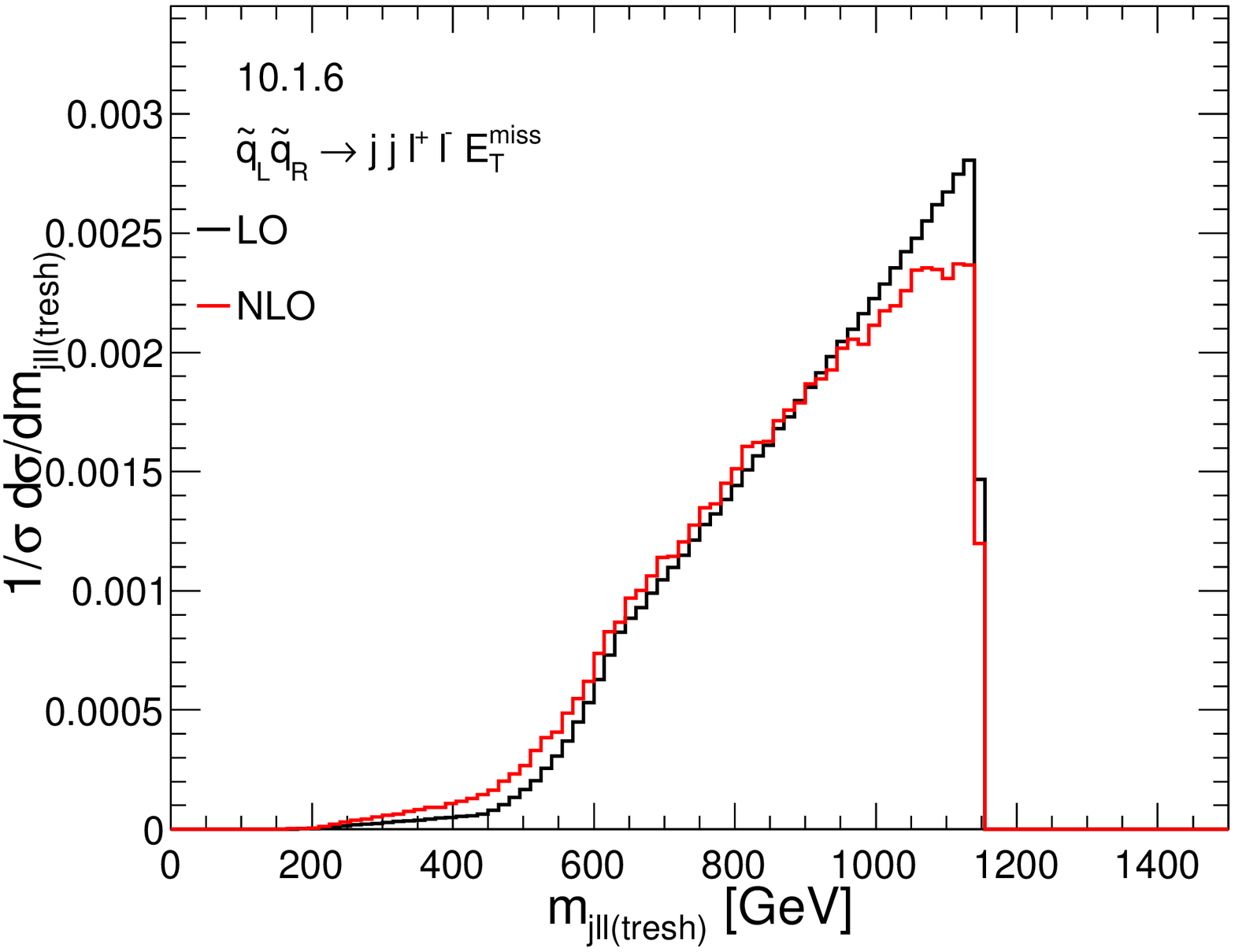}\\
\caption{LO (black) and NLO (red) normalized differential distributions for 10.1.6 in $\mjllow$, $\mjlhigh$, $\mjll$ and $\mjlltresh$ (from top left to bottom right) for combined production and decay.
}
\label{fig:combined_1016}
}
\FIGURE{
\includegraphics[width=.45\textwidth]{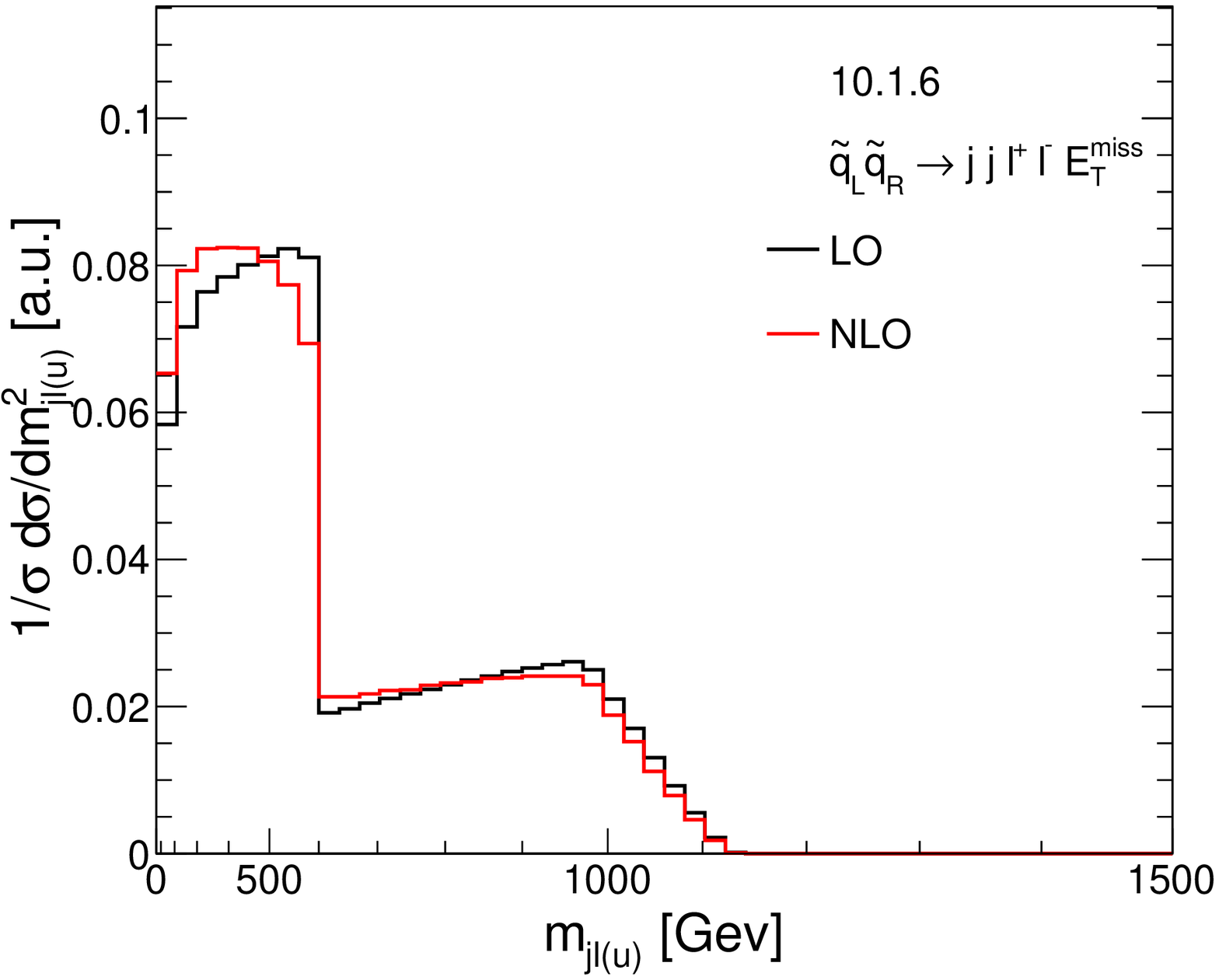}
\includegraphics[width=.45\textwidth]{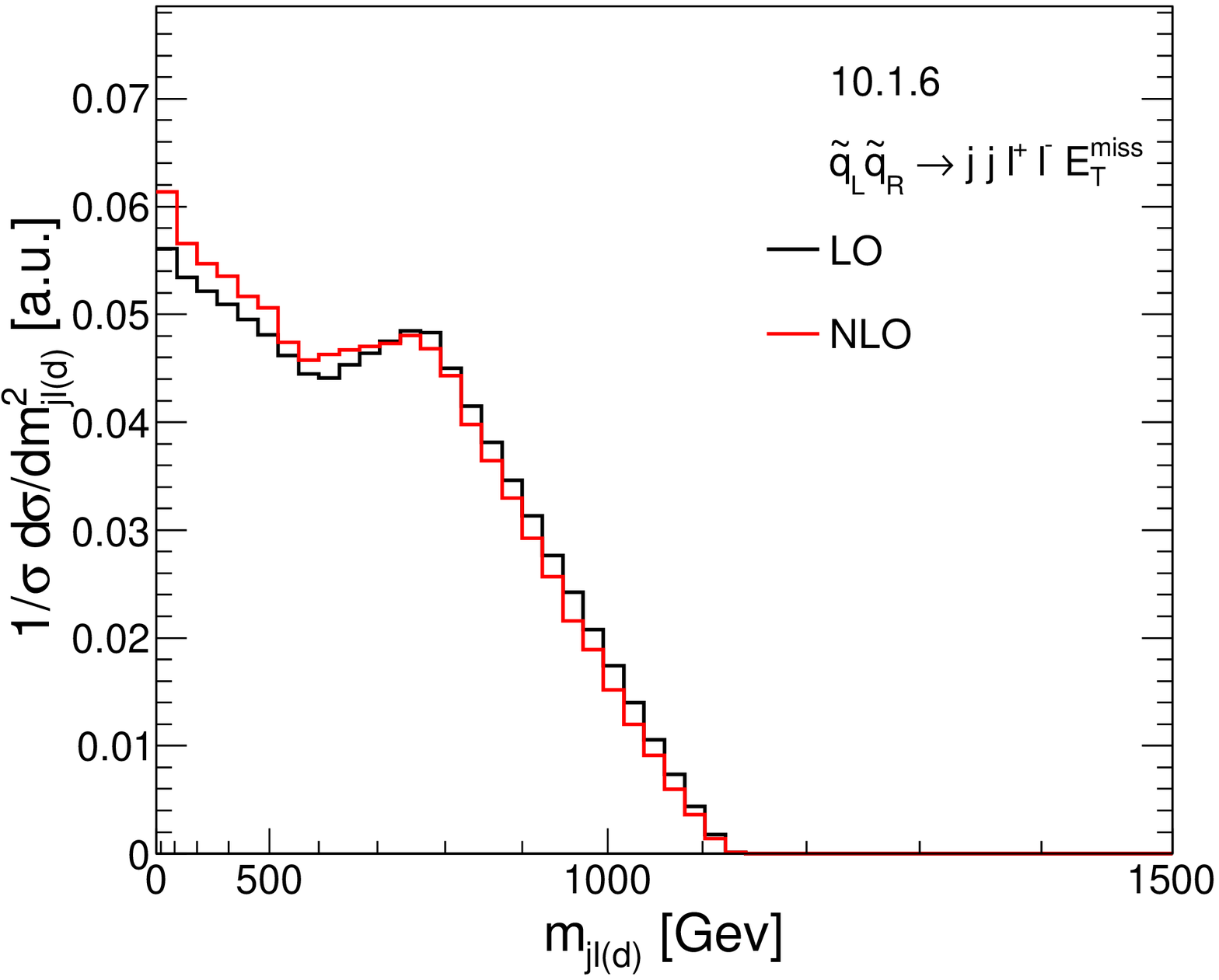}\\
\includegraphics[width=.45\textwidth]{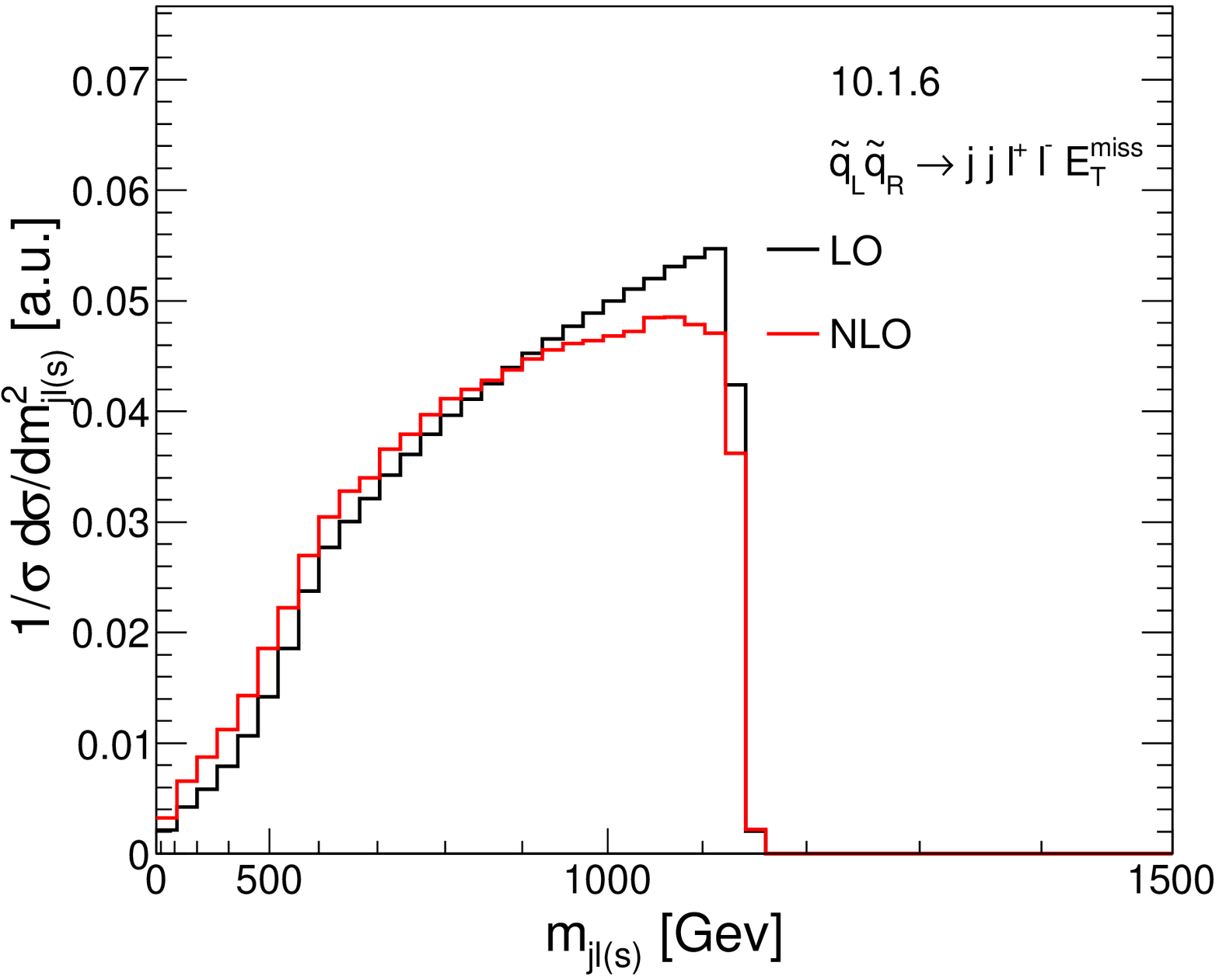}
\includegraphics[width=.45\textwidth]{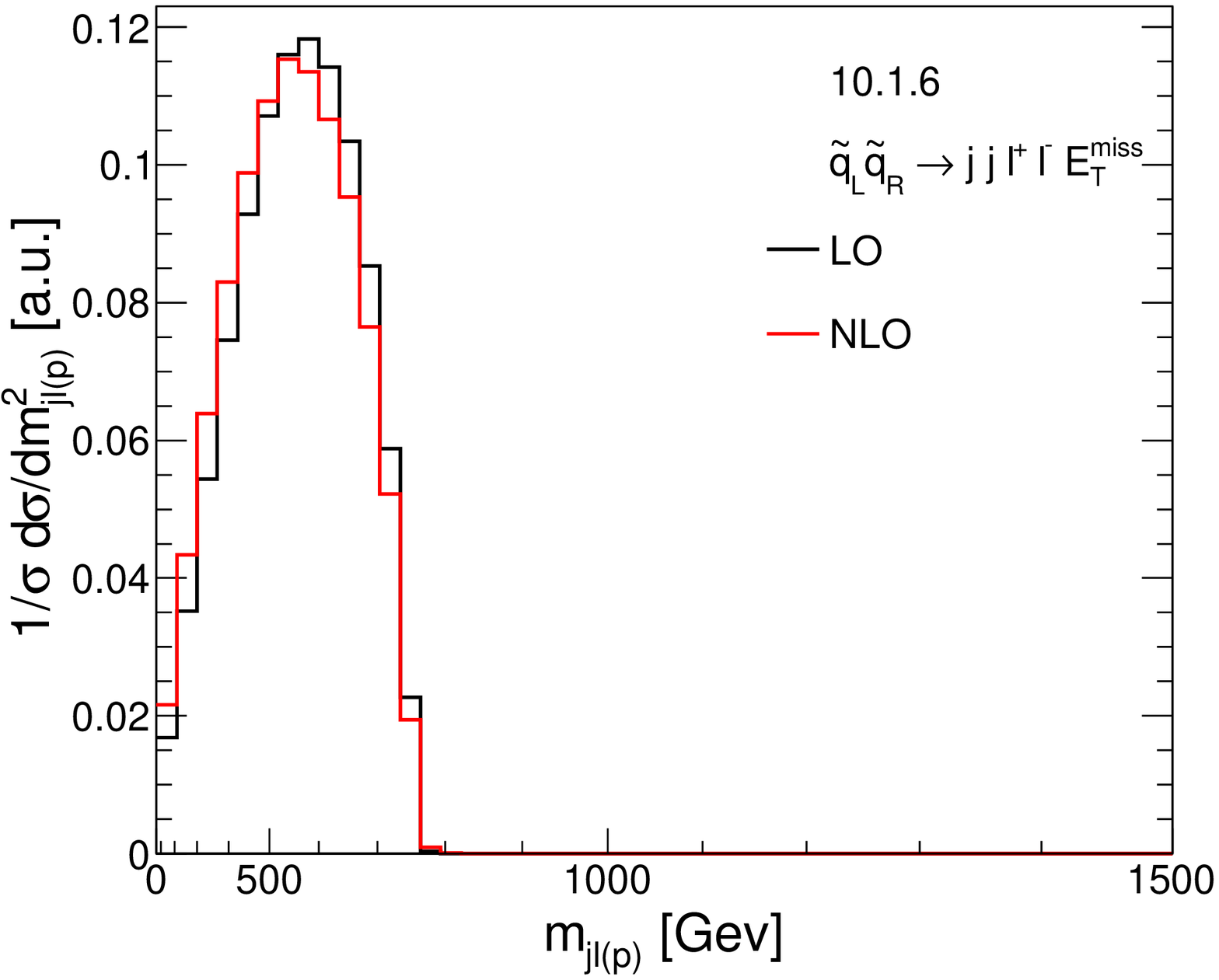}\\
\caption{LO (black) and NLO (red) normalized differential distributions for 10.1.6 in $\mjlu$, $\mjld$, $\mjls$ and $\mjlpp$ (from top left to bottom right) for combined production and decay shown with a quadratic scale. 
}
\label{fig:combined_1016_matchev}
}

\FIGURE{
\includegraphics[width=.45\textwidth]{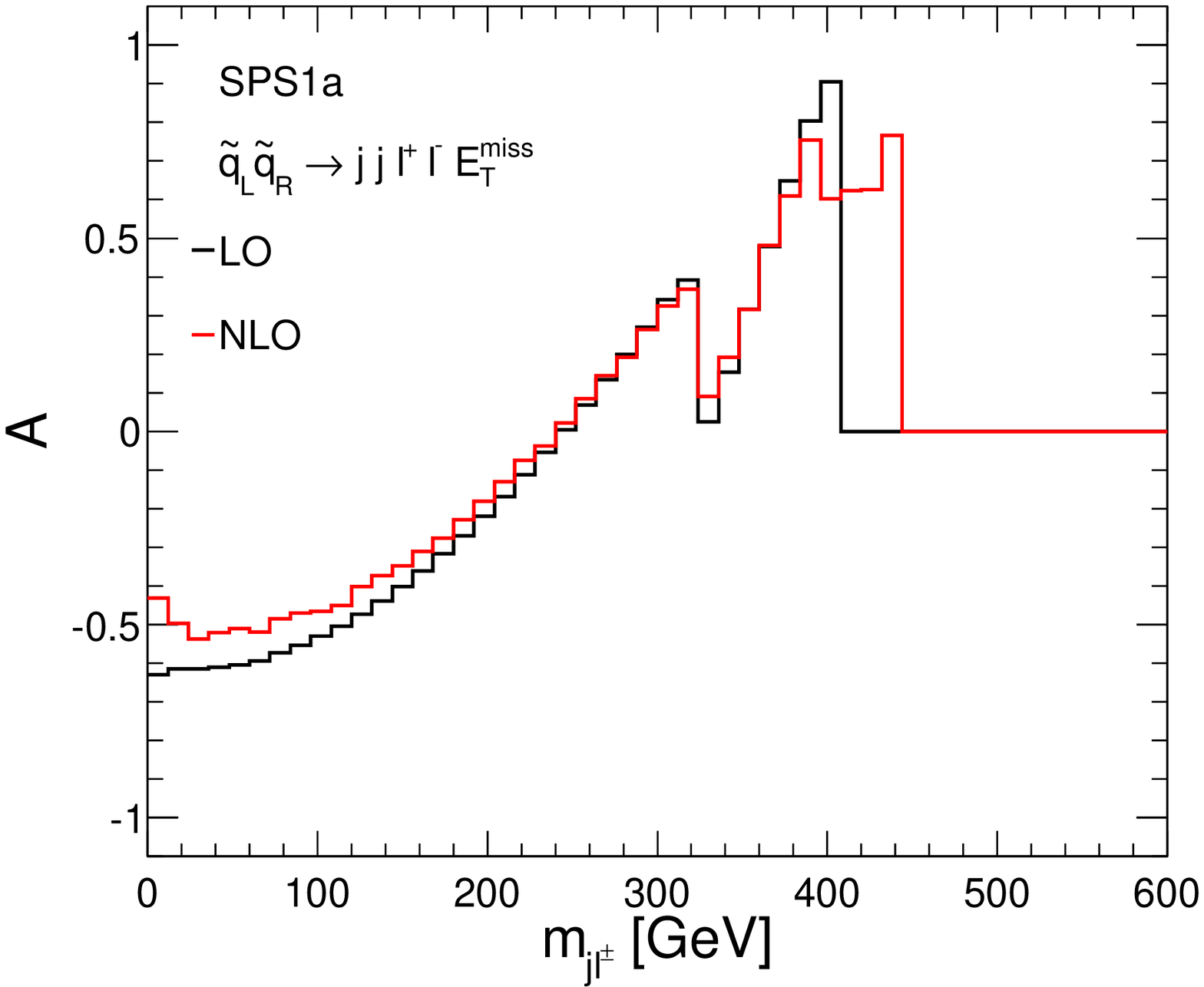}
\includegraphics[width=.45\textwidth]{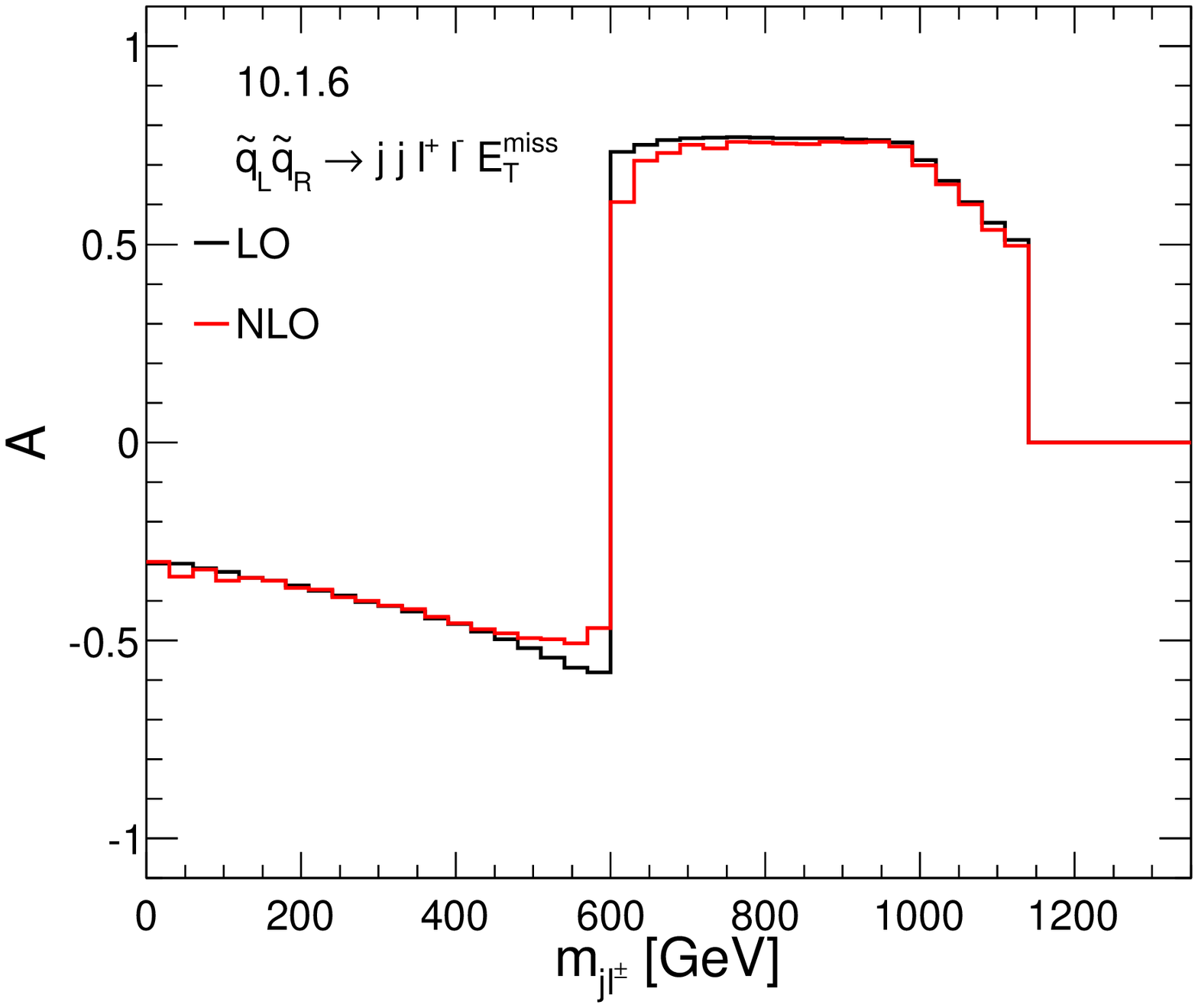}
\caption{LO (black) and NLO (red) normalized differential distributions for SPS1a (left) and 10.1.6 (right) in the asymmetry A for combined production and decay. 
}
\label{combined_A}
}

\clearpage

\subsubsection{Inclusive observables}
\label{sec:num_inc}
As discusssed in the previous sections, the experimental signature of two jets, two OS-SF leptons and missing energy provides essential information for future SUSY parameter determination. Additonally this signature can also be used for searches for supersymmetry. At the LHC, using this signature, cut-and-count searches have been performed in the analyses of the 7 and 8 \TeV runs \cite{Chatrchyan:2012te,Aad:2011xk} and will be performed also at 14 \TeV. Consequently, precise calculations of inclusive cross sections for the specific signal region used in the cut-and-count searches are necessary.

\TABULAR[t]{c|cc|cc||cc}{
\hline
   & $\boldsymbol{N^{(0)}_{2j+2l+\not E_T}}$ 
& $\boldsymbol{N^{(0),\text{cons. cuts}}_{2j+2l+\not E_T}}$ 
 & $ \boldsymbol{K_{N_{2j+2l+\not E_T}}}$ & $ \boldsymbol{K^{\text{cons. cuts}}_{N_{2j+2l+\not E_T}}}$ &
$\boldsymbol{K_{pp\to \sq{\vphantom{'}}_L\sq'_R}}$ & $\boldsymbol{K_{pp\to \sq\sq'}}$ 
\\
\hline
SPS1a		& $38.2$\fba	& $23.0$\fba & $1.36$	& $1.23$	& $1.34$	& $1.28$\\
\hline
10.1.6		& $0.628$\fba & $0.243$\fba	& $1.46$	& $1.39$ 	& $1.44$	& $1.41$\\ \hline}
{LO $N^{(0)}$ and NLO $N^{(0+1)}$ cross section
predictions and K-factors $K_{N}$ for the two benchmark scenarios SPS1a,
10.1.6 at a center of mass energy $\SqrtS=14~\TeV$ where the cuts of
\eqref{eq:cuts_combined} are applied. For comparison we also list the inclusive NLO production
K-factor $K_{pp\to \sq\sq'}$ and $K_{pp\to \sq_L\sq_R'}$.  
\label{tab:cmsrates}
}

First, we discuss the differences between predictions in our approximation and rescaling LO with a flat K-factor from NLO corrections to squark-squark production without including decays and cuts. Second, we look at the impact of the additional consistency cuts defined in \eqref{consistency}.  
In \tabref{tab:cmsrates} various integrated quantities at 14 \TeV for the parameter points SPS1a and 10.1.6 are listed. Starting from the first column on the left we display: the total cross section at LO $N^{(0)}_{2j+2l+\not E_T}$ in the signal region defined by the cuts of \eqref{eq:cuts_combined}, the total cross section at LO after consistency cuts $N^{(0),\text{cons. cuts}}_{2j+2l+\not E_T}$,
together with the corresponding K-factors $K_{N_{2j+2l+\not E_T}}$ and $ K^{\text{cons. cuts}}_{N_{2j+2l+\not E_T}}$. Furthermore, we list the K-factors for the production only including all $ \sq{\vphantom{'}}_L\sq'_R$ channels $K_{pp\to \sq{\vphantom{'}}_L\sq'_R}$ and including also all other chirality configurations $K_{pp\to \sq\sq'}$.

The difference between the K-factors including the cuts defining the signal region, $K_{N_{2j+2l+\not E_T}}$, and the K-factors for production of $\sq{\vphantom{'}}_L\sq_R$ pairs, $K_{pp\to \sq{\vphantom{'}}_L\sq_R}$, is small, namely $2\%$. This difference increases to $8\%$ for SPS1a and $5\%$ for 10.1.6 if the K-factor for just the production includes also the other chirality configurations, $K_{pp\to \sq\sq'}$. Thus, for the scenarios analyzed here, NLO corrections can be safely approximated rescaling LO predictions with the K-factor obtained for the production part, provided that only the contributing chirality configurations are included in the calculation of the K-factor.
This feature, however, cannot easily be generalized; for example,  as can be seen from \tabref{tab:cmsrates}, 
applying consistency cuts increases the  differences between these two approximations.

Consistency cuts are designed for the study of distributions relevant for parameter determination, as the ones discussed in the previous section. However, they decrease the cross sections, as can be seen comparing $N^{(0)}_{2j+2l+\not E_T}$ and $N^{(0),\text{cons. cuts}}_{2j+2l+\not E_T}$, without adding any obvious benefit in the context of searches. On the other hand, so far, we did not discuss the different normalization of the LO and NLO distributions shown in the previous sections. The values of $K^{\text{cons. cuts}}_{N_{2j+2l+\not E_T}}$ are exactly the ratios between the normalization of the LO and NLO results. These values are smaller than the K-factors obtained without consistency cuts. As discussed, such cuts are ideated to 
solve the jet combinatorical problem. Doing so, they also reduce (positive) contributions from real radiation of a gluon or a quark at NLO and consequently reduce the resulting K-factors. Thus, in contrast to the case without consistency cuts applied, fully differential factorizable corrections have to be taken into account for a precise estimation of NLO effects.

%
\pagebreak
\section{Conclusions}
\label{sec:conlusion}
In this paper we provided, for the first time, an analysis at NLO of the contribution from squark-squark production to the experimental signature  $2j+l^{+}l^{-}+\missingET(+X)$, taking into account decays and experimental cuts.
We focused on the impact of NLO corrections on invariant mass distributions that can be used, in case of discovery of supersymmetric particles, for parameter determination. We observe that general shapes, besides smoothing of edges and kinks and a shift towards smaller invariant masses, are not strongly altered. This seems to be an universal behaviour despite the strong dependence of shapes on the parameter region.

We also analyzed the impact of NLO corrections including decays on the predictions for cut-and-count strategies used in discovery searches. The predictions depend on the cuts applied and in general can be different from the result obtained rescaling LO predictions with flat K-factors obtained from the cross-section of just production without decays and cuts included. However, in particular cases, results obtained in this approximation and using our calculation can be in very good agreement, provided that only the contributing chirality configurations are included in the calculation of the flat K-factor.

In our framework, results can be easily extended including off-shell effects in the electroweak decay chain. Also different kinds of experimental signatures emerging from different decay chains can be analyzed easily. In general an analogue calculation of such signatures from the other production channels of squarks and gluinos is desirable.


\acknowledgments
This work was supported in part by the Research Executive Agency 
of the European Union under the 
Grant Agreement PITN-GA-2010-264564 (LHCPhenonet).
We acknowledge use of the computing resources at the Rechenzentrum Garching.
%

\bibliographystyle{JHEP}
\providecommand{\href}[2]{#2}\begingroup\raggedright
\endgroup

\end{document}